\documentclass[10pt]{elsarticle}

\journal{Astroparticle Physics}

\usepackage{geometry} 

\usepackage{graphicx}

\usepackage{lscape}
\usepackage[T1]{fontenc}
\usepackage{color}

\usepackage{sidecap}
\sidecaptionvpos{figure}{c} 

\usepackage{subfigure}   

\usepackage{multirow}
\usepackage{booktabs}

\usepackage{mdwlist}

\usepackage{amsmath,amssymb} 
\usepackage{placeins}
\usepackage{rotating}

\usepackage{wrapfig}

\usepackage{subfigure}

\usepackage{setspace}

\usepackage{comment}
\usepackage{hyperref}

\newcommand{\fix}[1]{\textbf{\textsc{\textcolor{red}{}}}}
\newcommand{\tfix}[1]{\textbf{\textsc{\textcolor{green}{}}}}
\newcommand{\kfix}[1]{\textbf{\textsc{\textcolor{blue}{}}}}

\newcommand{\fixed}[1]{}

\renewcommand{\vec}[1]{\textbf{#1}}

\newcommand{\be}{\begin{equation}} 
\newcommand{\ee}{\end{equation}}
\newcommand{\bea}{\begin{eqnarray}} 
\newcommand{\eea}{\end{eqnarray}}

\newcommand{\bc}{\begin{centering}}
\newcommand{\ec}{\end{centering}}

\newcommand{\bkgaf}{\begin{figure}} 
\newcommand{\ekgaf}{\end{figure}}

\newcommand{\BP}{\begin{quotation} \begin{em} \begin{small} \singlespacing}
\newcommand{\EP}{ \end{small}\end{em}\end{quotation} \doublespacing}

\newcommand{\lna}{$\langle\ln A\rangle$}
\newcommand{\s }{$S_{125}$}
\newcommand{\K}{$K_{70}$}

\usepackage{url}
\makeatletter
\def\url@leostyle{%
  \@ifundefined{selectfont}{\def\UrlFont{\sf}}{\def\UrlFont{\small\ttfamily}}}
\makeatother
\title{Cosmic Ray Composition and Energy Spectrum from 1--30 PeV Using the 40-String Configuration of IceTop and IceCube}

\author[Madison]{R.~Abbasi}
\author[Gent]{Y.~Abdou}
\author[Zeuthen]{M.~Ackermann}
\author[Christchurch]{J.~Adams}
\author[Geneva]{J.~A.~Aguilar}
\author[Madison]{M.~Ahlers}
\author[Berlin]{D.~Altmann}
\author[Madison]{K.~Andeen}
\author[Madison]{J.~Auffenberg}
\author[Bartol]{X.~Bai\fnref{SouthDakota}}
% ...
\address[Berlin]{Institut f\"ur Physik, Humboldt-Universit\"at zu Berlin, D-12489 Berlin, Germany}
% ...
\address[Christchurch]{Dept.~of Physics and Astronomy, University of Canterbury, Private Bag 4800, Christchurch, New Zealand}
% ...
\address[Geneva]{D\'epartement de physique nucl\'eaire et corpusculaire, Universit\'e de Gen\`eve, CH-1211 Gen\`eve, Switzerland}
\address[Gent]{Dept.~of Physics and Astronomy, University of Gent, B-9000 Gent, Belgium}
% ...
\address[Madison]{Dept.~of Physics, University of Wisconsin, Madison, WI 53706, USA}
% ...
\address[Bartol]{Bartol Research Institute and Department of Physics and Astronomy, University of Delaware, Newark, DE 19716, USA}
% ...
\address[Zeuthen]{DESY, D-15735 Zeuthen, Germany}
\fntext[SouthDakota]{Physics Department, South Dakota School of Mines and Technology, Rapid City, SD 57701, USA}
\fntext[LosAlamos]{Los Alamos National Laboratory, Los Alamos, NM 87545, USA}
\fntext[Bari]{also Sezione INFN, Dipartimento di Fisica, I-70126, Bari, Italy}
\fntext[Goddard]{NASA Goddard Space Flight Center, Greenbelt, MD 20771, USA}

\begin{document}


\begin{abstract}
The mass composition of high energy cosmic rays depends
on their production, acceleration, and propagation. The study of cosmic ray composition can therefore reveal hints of the origin of these particles. At the South Pole, the IceCube Neutrino Observatory is capable of measuring two components of cosmic ray air showers in coincidence:
the electromagnetic component at high altitude (2835~m) using the IceTop surface array, and the muonic component above $\sim$1 TeV using the IceCube array.  This unique detector arrangement provides an opportunity for precision measurements of the cosmic ray energy spectrum and composition in the region of the knee and beyond.
We present the results of a neural network analysis technique to study the cosmic ray composition and the energy spectrum from 1~PeV to~30~PeV using data recorded using the 40-string/40-station configuration of the IceCube Neutrino Observatory.   

\end{abstract}

\maketitle


%

\section{Introduction}

The flux of cosmic rays at Earth is known to follow a steep power-law spectrum over a large energy range.  The index of this spectrum is approximately constant at energies lower than about 3~PeV, where the spectrum steepens in a feature known as the ``knee''.  A further kink in the spectrum, known as the ``ankle'', occurs around 1 EeV where the spectrum becomes flatter again.  The origins of these spectral changes are still uncertain.
Currently, the most popular model 
predicts cosmic ray acceleration in shock fronts via the first order Fermi mechanism \cite{Bell:1978}.  
More specifically, at energies up to~$\sim10^{17}$~eV, 
the source of this acceleration mechanism is often attributed to supernova remnants; a cut-off energy which depends upon nuclear charge ($Z$) of the particle accelerated at the source could be responsible for a mass-dependent knee; the ankle is then attributed to cosmic rays from extragalactic sources such as gamma-ray bursts or active galactic nuclei \cite{Horandel:2009, Torres:2004, Stanev:2007}.  

Various underlying source, acceleration, and propagation models, though tuned to predict similar energy spectra, differ considerably in energy-dependent composition in the region between the knee and the ankle \cite{Horandel:2004}.  This dependence on primary mass implies that a precise measurement of the composition of cosmic rays would provide important clues as to the origins of these particles.  However, due to the decreasing flux of cosmic rays with increasing energy, 
measurements with high statistics are difficult to collect with satellite or balloon-based detectors above 100~TeV. 
Ground-based detectors can observe cosmic ray air showers above 100~TeV indirectly with high statistics.
Indirect measurements of composition involve a close examination of the extensive air shower produced by a primary cosmic ray particle colliding with Earth's atmosphere.  By using information from more than one component of the shower, such as the electromagnetic and muonic components, the energy and composition can be obtained for primary particles with much higher energies than those currently measurable through direct detection techniques.  

At the South Pole, the IceCube Neutrino Observatory is sensitive to air showers with energies 
from below the knee to
the ankle region of the energy spectrum. This region covers the predicted transition from galactic to extragalactic cosmic ray sources.  The IceTop surface array measures a combination of the electromagnetic and the low energy muonic components of the cosmic ray shower, while the IceCube array measures the bundle of high-energy muons ($>$300 GeV) deep under the surface of the ice.  In the following analysis, data from these two components are combined for a coincident composition measurement.

\section{The IceCube Neutrino Observatory}
\label{sec:intro}
The IceCube Neutrino Observatory consists of two parts: 
IceTop, a surface air shower array \cite{IceTopDetectorPaper:2012}, and IceCube, 
a muon and neutrino telescope installed deep in the ice.  These detectors are the successors to the
SPASE \cite{Dickinson:2000} and AMANDA \cite{Ahrens:2002} experiments.
Each array is comprised of light sensors called Digital Optical Modules (DOMs) \cite{Abbasi:2009}, which
detect Cherenkov photons emitted by relativistic charged particles passing through ice.  
Each DOM is a spherical, pressure-resistant glass shell containing a 
25~cm diameter Hamamatsu photomultiplier tube (PMT),
a mu-metal grid for magnetic shielding of the PMT, and electronics for operation and control of the 
PMT as well as amplification, digitization, filtering, and calibration.

\begin{figure}[htbp]
\bc
\subfigure[]{\includegraphics[height=0.65\textwidth]{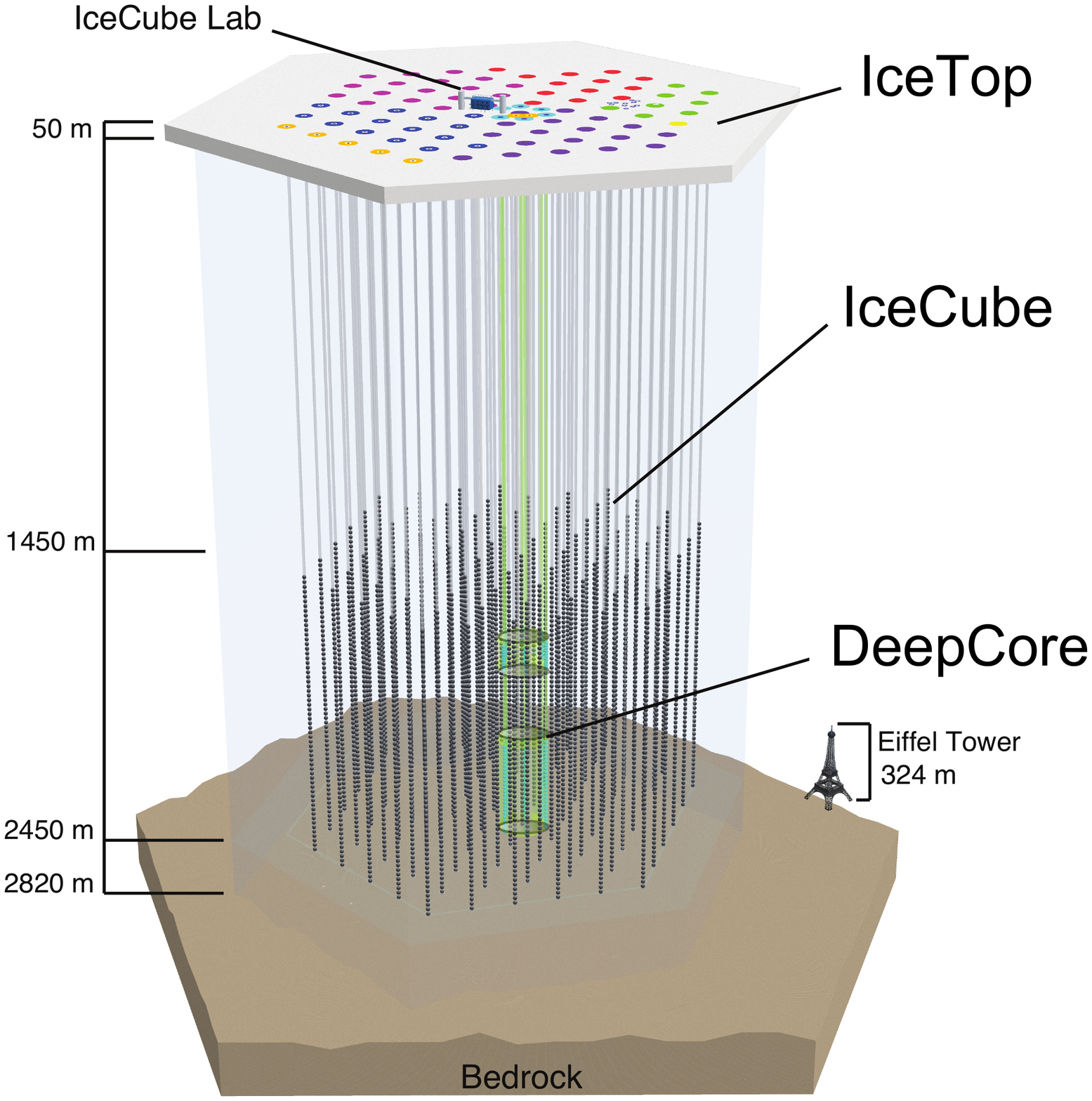}}
\subfigure[]{\includegraphics[height=0.65\textwidth]{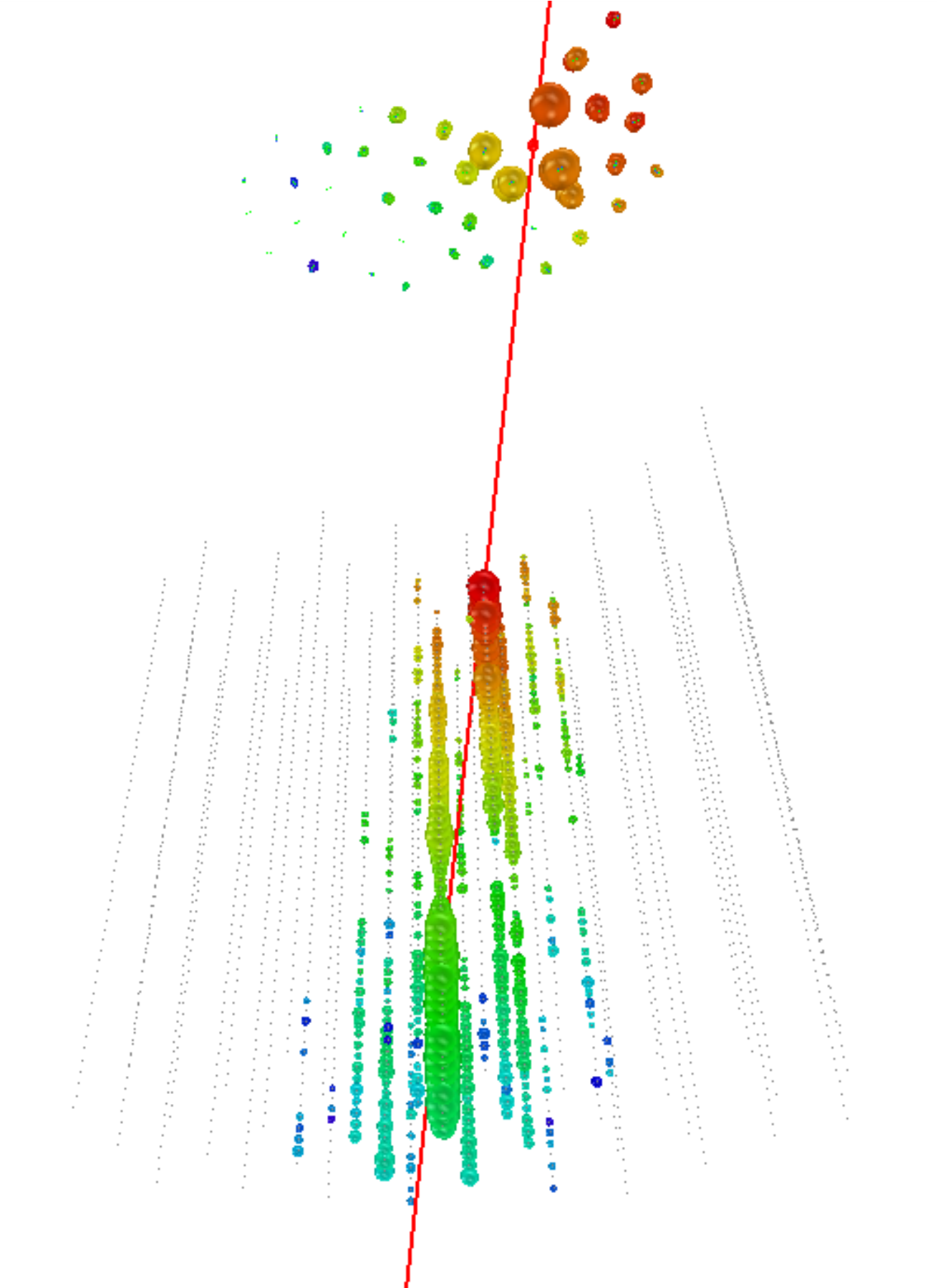}}
\caption[Event Display and Detector Setup]{\footnotesize {\it Left}: A schematic of the final IceCube Neutrino Observatory layout, completed in 2011.  In 2008, only the IceTop and IceCube arrays existed, though a low-energy in-fill array called DeepCore has since been added. {\it Right}: A coincident event from the IceTop/IceCube 40-string configuration of 2008.  The colors represent the timing of the hits (red is earliest, blue is latest), and the size of the sphere around a DOM represents the amplitude of light seen by that DOM.  In this large event, a big dust layer can clearly be distinguished as a ``waist'' in the amplitudes just over half-way down the IceCube array.  }
\label{iti3_final}
\ec
\end{figure}

In IceCube, DOMs are frozen into the ice 
along strings which are placed in a 125~m triangular grid formation.  The DOM's are vertically spaced 17~m apart, at depths from 1450~m to 2450~m below the surface, as shown in Fig.~\ref{iti3_final}(a).
The direction of muons (either from cosmic ray air showers above the surface, or neutrino interactions within the ice or bedrock) can be reconstructed from the pattern (position and timing) of hit DOMs.  This feature is demonstrated in Fig.~\ref{iti3_final}(b), which shows a coincident IceTop/IceCube event from the 40-string/40-station configuration of 2008.  
The minimum energy at the surface required for a muon to penetrate through the ice to the top of the IceCube array is around 300~GeV, while the threshold for the muon to pass through the whole detector volume is around 500~GeV.  

Each IceCube string is associated with an IceTop ``station'', which consists of two cylindrical ice Cherenkov tanks separated by 10~m.  
Each tank has an inner diameter of 1.82~m, is 1.3~m high, and contains two down-facing DOMs, with center-to-center spacing of 58~cm, frozen in optically clear ice 90~cm in depth.  The PMT in each DOM has an adjustable gain; to increase the dynamic range of the detector, one DOM in each tank is operated at low gain (LG) while its partner is operated at high gain (HG).  The tanks are lined with a diffusely reflective coating of either Tyvek or, in most cases, zirconium fused polyethylene, and the surface of the ice is covered with perlite.
IceTop measures mainly the electromagnetic component of incoming cosmic ray air showers.  
The signal response of a particle passing through an IceTop tank is measured in photoelectrons (PEs); however, each tank must be calibrated to obtain a uniform measurement.  Therefore, the amount of charge deposited by mainly vertical 1 -- 10~GeV surface muons 
passing through a given tank is defined in terms of a common unit, called a ``vertical equivalent muon", or VEM.  A calibration process is then performed by comparing the charge spectra of each DOM -- which shows a clear muon peak -- in data and simulation.  
The altitude of the surface array is 2835~m, which corresponds to an atmospheric depth of $\sim$680~g/cm$^2$.
For showers with tens of PeV energies, this is just below the proton shower maximum; therefore, an excellent energy resolution is expected.

Construction of the IceCube Neutrino Observatory 
began at the geographic South Pole in the 2004--05 austral summer
\cite{Achterberg:2006} and was completed during the 2010--11 austral summer.
This work used the configuration of the detectors operational from  April 2008 to May 2009, which consisted of 40 IceCube strings (2400 DOMs) and 40 IceTop stations (160 DOMs).  
Figure~\ref{iti3_final}(a) shows the final detector configuration of 86 strings, while
Fig.~\ref{iti3_final}(b) is an example data event from the 40-string
configuration studied in this work.

\section{Data and Simulation}
\subsection{Data} 
This analysis uses data from August 2008, when the detector was in its 40-string/40-station configuration, for an overall detector livetime of 29.78 days. 
A study of the relationship between the electromagnetic and muonic air shower observables (\K{}~and \s{}, as discussed in Section~\ref{sec:reco}) throughout the full year of data from the 2008 configuration revealed
significant seasonal variations in the relationship between the observables used for this analysis \cite{Andeen:thesis}.  
In particular, the effective temperature of the atmosphere (which is stable from day to day but varies dramatically between summer and winter) 
affects the production of muons in the upper atmosphere, changing the measured \K{}.
Limiting the data set to August 2008, a mid-winter month which corresponds to our simulated data (described in Section~\ref{sec:sim}), 
effectively mitigates these fluctuations.

Local coincidence (LC) and trigger requirements are the first steps in the data acquisition system of the detectors \cite{Abbasi:2009}.  
The LC requirement is satisfied in IceCube when two neighboring or next-to-nearest-neighboring DOMs on a string pass a signal threshold of 0.25~photoelectrons (PE) within 1~$\mu$s.
The LC requirement is satisfied in IceTop when either the HG DOMs in both tanks of a station, or a HG DOM and the LG DOM from the neighboring tank in the station have passed the signal threshold of about 20 PE within 1~$\mu$s.
An event has triggered a detector when a certain number of DOMs record LC signals within a 5~$\mu$s sliding time window: eight DOMs for IceCube, and six DOMs for IceTop.
Additionally, a ``coincident event'' is one where both detectors have triggered.

\subsection{Simulation}
\label{sec:sim}
Monte Carlo simulated events were produced for this analysis using the \textsc{CORSIKA} air shower generator \cite{Heck:2010} with the SIBYLL-2.1/FLUKA-2008 hadronic interaction models \cite{Ahn:2009, FLUKA:2006}. Five particle species (proton, helium, oxygen, silicon, and iron) were generated with an $E^{-1}$ spectrum from 10~TeV to 50~PeV.  The showers were generated uniformly over all azimuths and with a sin($\theta$)cos($\theta$) distribution in the zenith range from $0^{\circ}$ to $65^{\circ}$, which is steeper than that reached by reconstructible coincident events.  For each species, 3000 showers were simulated per third of a decade in energy.  Each shower was oversampled 100 times over a circle with radius of 1200~m leading to a total of 16.5 million generated events.  The atmospheric model chosen corresponds to the austral winter months at the South Pole.  For initial comparison with experimental data, the simulation is reweighted, independent of primary mass, to an $E^{-2.7}$ spectrum at energies below a knee at 3~PeV and an $E^{-3.0}$ spectrum above the knee.
The IceTop detector is simulated using parameterizations of the tank response obtained from a complete \textsc{Geant}4 \cite{GEANT4:2003, GEANT4:2006} simulation.
The muons are then propagated through the ice to the depth of the IceCube array using the muon propagator \textsc{MuonMonteCarlo} \cite{Chirkin:2004} which accounts for continuous and stochastic energy losses.  The Cherenkov photons from the muon bundles passing through the volume of the array are simulated using the software package \textsc{Photonics} \cite{Lundberg:2007}  which models the full structure of the ice properties according to the standard ice model used for the 40-string configuration of IceCube (known as the AHA ice model) {\cite{IceModel:2006}.  This is followed by a simulation of each aspect of the DOM electronics and the trigger.  The DOM signals are then processed in the same way as experimental data.  

In IceCube, the parameterization for the light yield from muon bundles in the \textsc{Photonics} software has recently been improved.  This improvement revealed an energy-dependent offset between our data and our standard simulation.  Thus, a small set of simulation was generated using the new software and this discrepancy was parameterized using a comparison of the two simulation sets.  This parameterization has been applied as a correction to the experimental data from this point forward.

\section{Reconstruction Algorithms}
\label{sec:reco}

\subsection{Reconstruction with IceTop}
Data collected by IceTop are analyzed to reconstruct the core position, 
direction and size of a cosmic ray air shower \cite{Klepser:thesis, Kislat:2011}. 
These parameters are determined by 
comparing the detected signal locations, charges, and times from hit stations
(as well as the locations of unhit stations) to what is expected from a 
cosmic ray air shower, according to a likelihood function:
\be
\mathcal{L} = \mathcal{L}_s + \mathcal{L}_0 + \mathcal{L}_t.
\label{eqn:llh}
\ee
where $\mathcal{L}_s$ is the likelihood of signal \emph{charges} from a lateral distribution function, $\mathcal{L}_0$ is the likelihood of \emph{unhit} stations, and $\mathcal{L}_t$ is the  likelihood of signal \emph{times} from an expected timing profile.  
All three are described in detail in Ref.~\cite{Kislat:2011}, but
a brief overview is given here.

The first term, $\mathcal{L}_s$, represents the likelihood of the array responding with 
measured signals in hit tanks $S_i$, given a set of expected signals for each 
of those hit tanks $S_i^{\mathrm{fit}}$.  Signals are assumed to have normal distributions
in log$_{10}$($S_i$), and are expressed in units of VEM (as discussed in 
Section~\ref{sec:intro}).

The signal expectation value, $S_i^{\mathrm{fit}}$, is a function of the 
perpendicular distance from the shower axis, $r$, and is given by a 
lateral distribution function which has been derived from simulations:
\be
S(r) = S_{\mathrm{ref}}\cdot\left(\frac{r}{R_{\mathrm{ref}}}\right)^{-\beta-\kappa \log_{10}\left(\frac{r}{R_{\mathrm{ref}}}\right)},
\label{eqn:s125}
\ee
where $\kappa$ is a constant optimized from simulations, and $\beta$ is fit as a free parameter to each
event.
The shower size, $S_{\mathrm{ref}}$, is the signal expectation at a reference distance, $R_{\mathrm{ref}}$, from the shower axis.  
The reference distance is chosen as the point in the fit which is the most stable or robust: multiple studies of the fit stability led to a choice of 125~m for this quantity \cite{Kislat:2011}.  
Figure~\ref{IT40_event} shows an example of the distribution of charges for one shower from the 2008 IceTop configuration with 40 stations. 
The superimposed curve is the lateral distribution function of Eq.~\ref{eqn:s125}
with the best-fit \s{} for this particular event.

\begin{figure}[htbp]
\begin{centering}
\subfigure[]{
\includegraphics[height=.45\textwidth,angle=90]{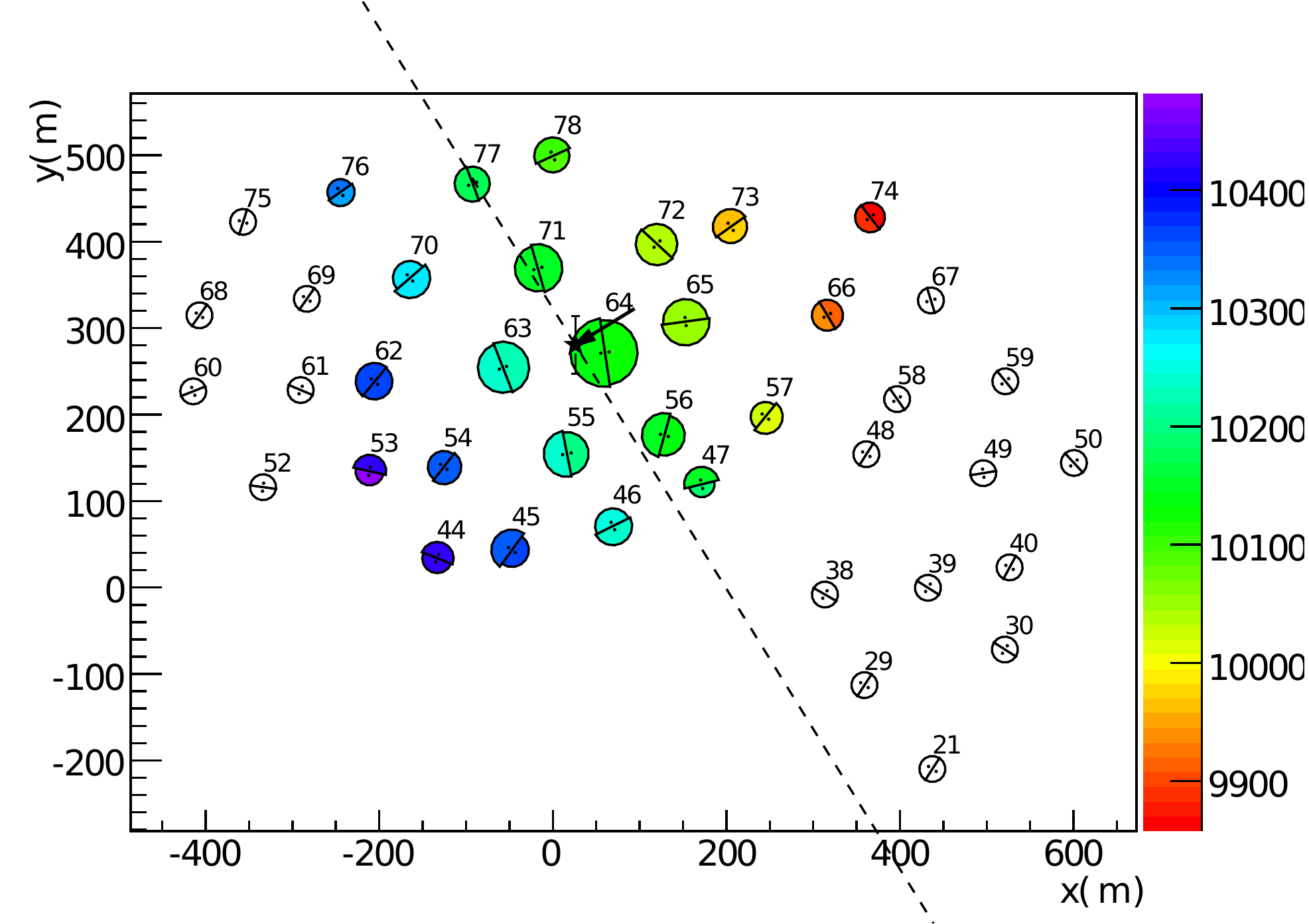}
\label{Tom_event_display}
}
\subfigure[]{
\includegraphics[height=.3\textwidth]{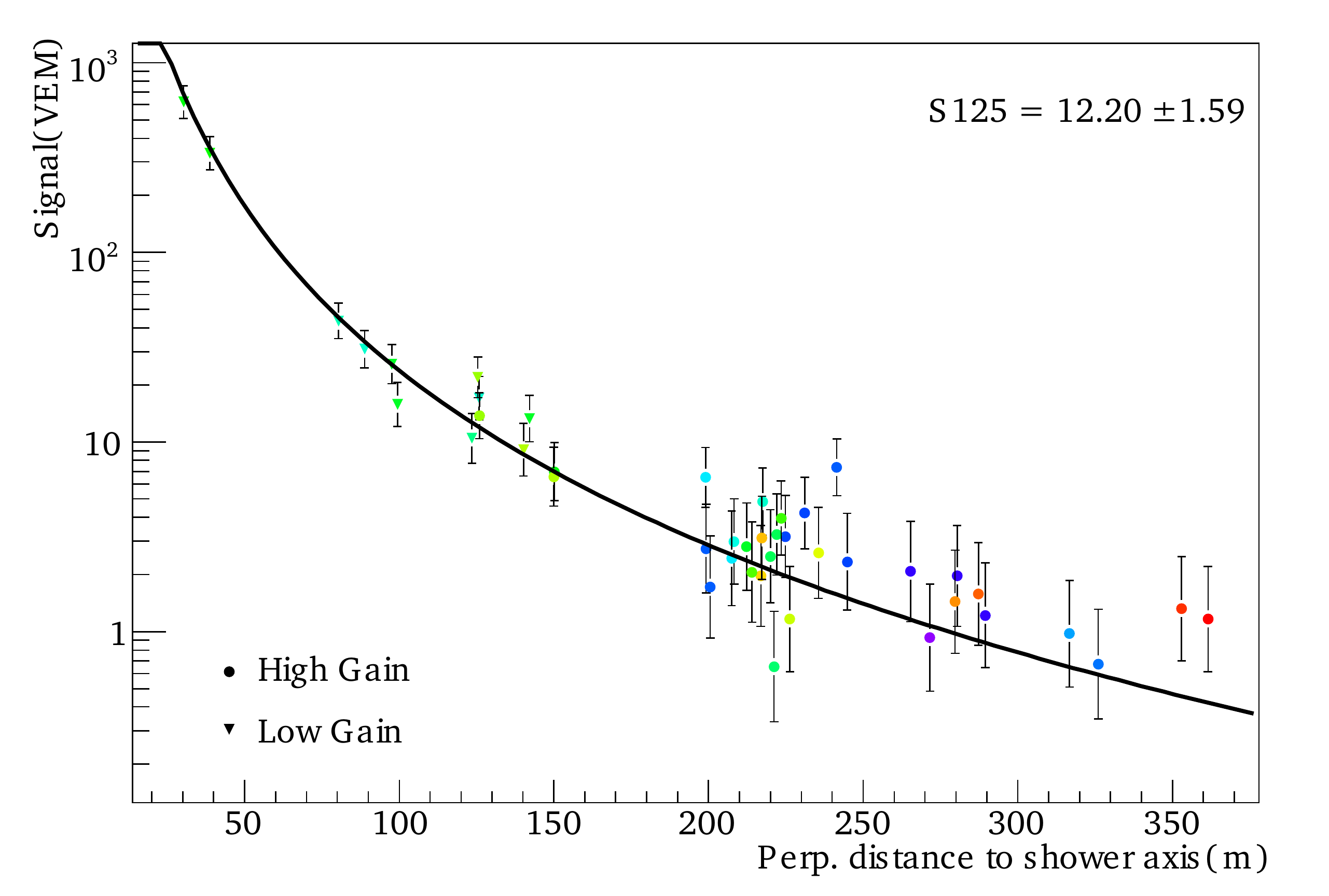}
\label{Tom_lateral_dist}
}
\caption[IceTop Event Display with Signal and Curvature Fits]
{\footnotesize An IceTop event from September 11, 2008.  {\it Left}: The core location 
is denoted with a star (near station 64) and the direction 
of the shower ($\theta \approx 15^{\circ}$, $\phi \approx 32^{\circ}$) is 
indicated by the arrow, and the dotted line represents the plane of the incoming shower front.  
The timing of the hits is represented by the colors, from red (earlier) to blue (later).  
The amount of signal observed by each {\it tank} in a given station is 
denoted by the size of the hemisphere closest to it.  
{\it Right}: The lateral distribution of these hits,
color-coded for timing in the same 
scheme as for the event display.
The expected lateral distribution of signals is shown in black;
\s{}~is this function evaluated at 125~m from the core of the shower; in this case, $12.2~VEM$.
}
\label{IT40_event}
\end{centering}
\end{figure}

The second likelihood term in Eq.~(\ref{eqn:llh}) 
accounts for all stations $j$ with the tanks ($j,A$) and ($j,B$) that did not trigger:  
\be
\mathcal{L}_0 = \sum_j \mbox{log$_{10}$} \left( 1 - P_{(j,A)}^{\mathrm{hit}}  P_{(j,B)}^{\mathrm{hit}}    \right),
\label{eqn:llh2}
\ee
where $P_{j,A(B)}^{\mathrm{hit}}$ is the probability that the ``A'' tank (or the ``B'' tank) within station $j$
did not trigger, for a given signal expectation value $S^{\mathrm{fit}}$.

The third term of Eq.~(\ref{eqn:llh}) describes the probability of measuring 
the observed signal arrival times, $t_i$, given expected signal
arrival times, $t_i^{\mathrm{fit}}$.  Each arrival time is assumed to be normally-distributed with
standard deviations $\sigma_{ti}(r_i)$, depending on the radial distance, $r_i$,  
of the tank to the shower core.

Studies of experimental data reveal that the shower front is not a flat plane.
Rather, the expected arrival times of the signals can be described as 
those expected from a flat plane, modified by a ``curvature term'' which is 
symmetric around the shower axis and a function of radial distance. 
The expected signal arrival time at a tank with 
position \vec{x} is thus parametrized as:
\be
t_{i}^{\mathrm{fit}} = t(\vec{x}) = t_0 + \frac{1}{c}(\vec{x$_c$} - \vec{x}) \cdot \vec{n} 
                + ar^2 + b\left(\exp\left(-\frac{r^2}{2\sigma ^2}  \right) -1\right).
\label{eqn:timedist}
\ee
Here, $t_{0}$ is the arrival time of the shower core at the surface of the ice 
(which is the surveyed level of the IceTop tanks), \vec{x$_c$} is 
the position of the shower core on the ground and \vec{n} is the 
unit vector in the arrival direction of the shower. 
The last two terms in Eq.~\ref{eqn:timedist} 
(the parabola and the Gaussian) 
together are the ``curvature term''.
The functional form of the curvature term and the 
values of the parameters $a$, $b$, and $\sigma$ 
were obtained from a study of the time residuals 
in experimental data.  

Thus, in this likelihood function there are seven total possible free parameters: \s{},
$\beta$, and the track's core position ($x_c$,$y_c$) and direction and core time ($\theta$, $\phi$, $t_0$).

\subsubsection{IceTop Fit Procedure.}
\label{sec:it_fit_proc}
Events in IceTop are initially reconstructed with two fast ``first guess''
algorithms.
First, the core position is estimated
using the center of gravity (COG) of the tanks which were hit, weighted with 
the square root of the charges:
\be
x_{COG} = \frac{\sum_i \sqrt{S_i}x_i}{\sum_i \sqrt{S_i}}.
\ee
Then, the direction of the shower is estimated 
by fitting a plane (without curvature terms) to the arrival times
of the signals.  
Both first guess core position and track direction together provide the seed track for
the more advanced lateral distribution fit involving all three likelihood terms
described in the previous section.

Multiple lateral distribution fits are performed in succession for best results.  
At each step a full minimization is performed.  
First, the shower direction is fixed and only the core position, shower size, 
and $\beta$ are allowed to vary as free parameters.  Next, since 
Eq.~(\ref{eqn:s125}) can exhibit unphysical behavior at small values of $r$, 
a second fit is made (again with shower direction fixed) in which 
all tanks closer than 11~m to the reconstructed shower core are 
excluded from the fit.  
If an 11~m circle around the {\it new} core location includes tanks which were
not excluded before (which happens rarely), they are added to the list of excluded tanks and
this second step is repeated until no new tanks are added to the list.
In a third step, the track direction is allowed to vary, as well as 
the core location and shower variables with smaller stepsizes. 
Finally, the track direction is fixed again 
and a final fine-tuning of just the core location and shower variables (with small stepsizes) 
is performed.

\subsection{Reconstruction with IceCube}
\label{sec:K70}
Bundles of muons that 
reach the IceCube detector
can have muon
multiplicities ranging from 1 to more than 1000 at the highest energies, and an
average lateral size of 20-30 meters.
When these muons pass through the ice at a speed greater than the
speed of light in the medium, they produce Cherenkov photons which can be detected
by the DOMs of the IceCube array.  Many techniques for reconstructing
the direction and energy of single particle tracks exist, which vary track parameters until the likelihood
of producing the measured amplitudes and arrival times of photons
at the DOMs is maximized \cite{Ahrens:2004a}.
But when the source of the photons is a muon bundle rather than
a single muon track, reconstruction can be improved with a likelihood
function specific to muon bundles,
which was developed using the
prototype detectors SPASE-2 and AMANDA-B10 \cite{Rawlins:thesis, Ahrens:2004}
and has been updated for use with IceCube.

\begin{figure}[tbp]
\bc
\subfigure[]{
\includegraphics[height=0.48\textwidth]{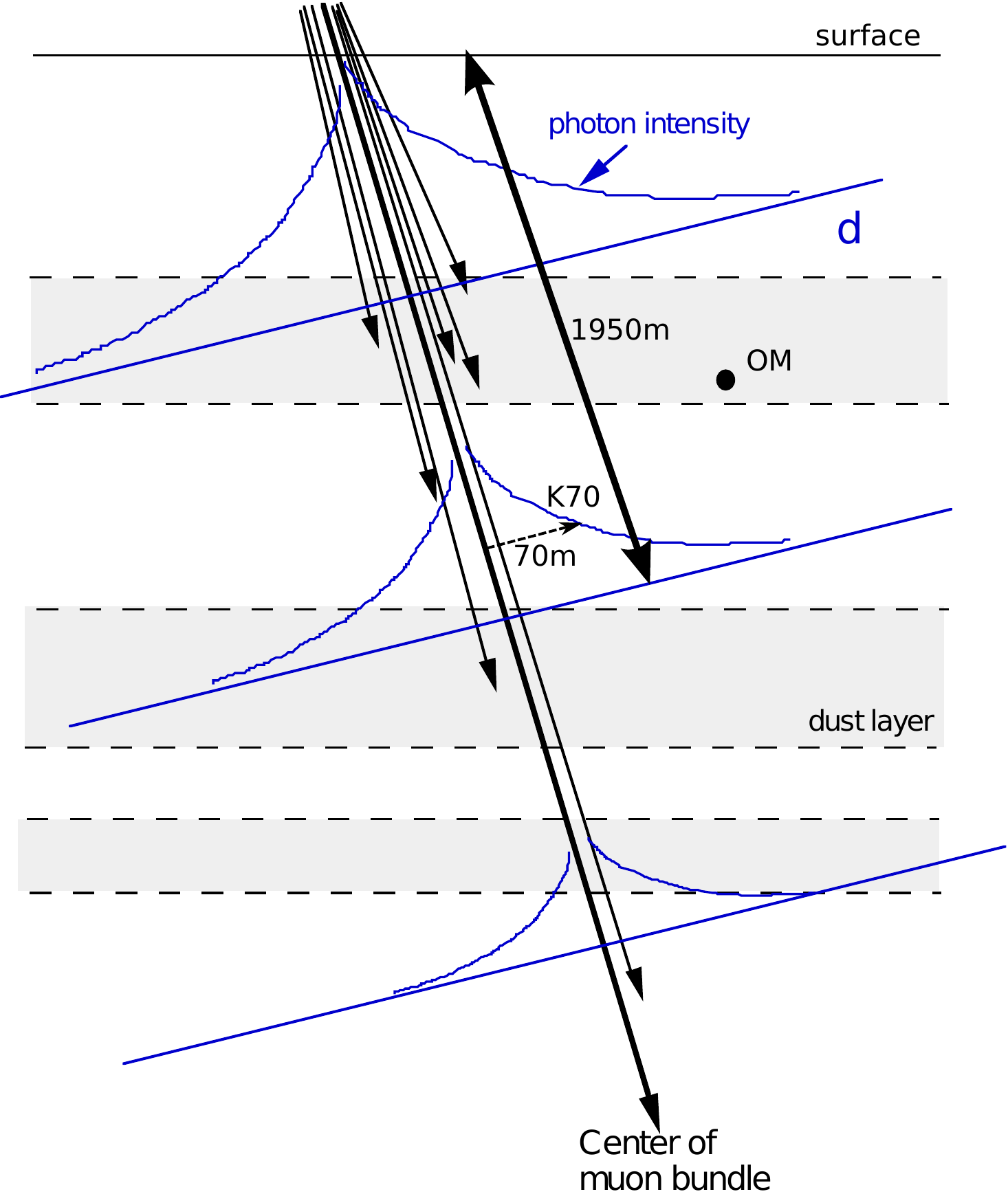}
\label{kath_adc_concept}
}
\subfigure[]{
\includegraphics[height=0.45\textwidth]{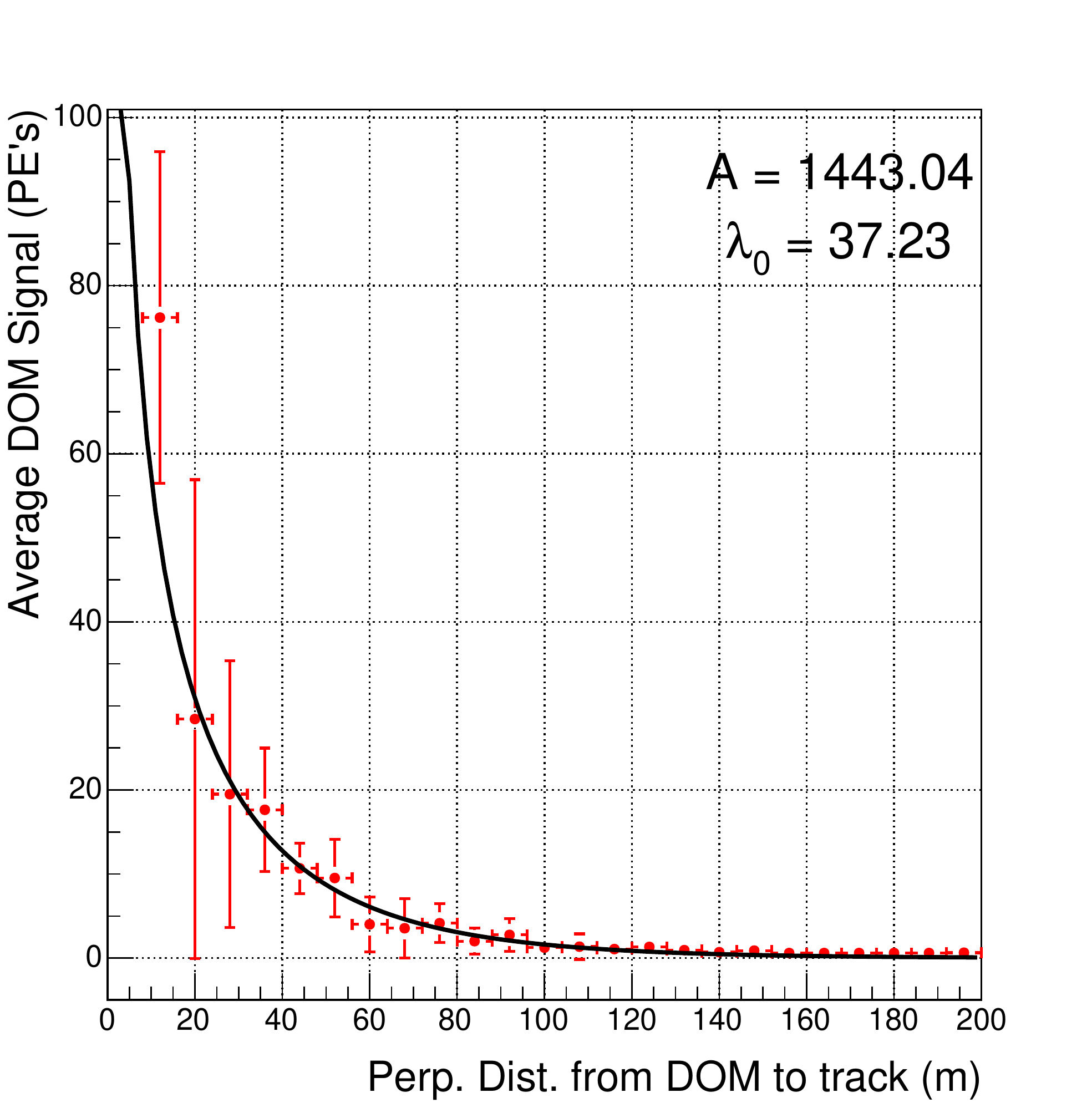}
\label{k-fit}
}
\caption[Cartoon of Muon Bundle in the Ice with Lateral Distribution of Light Intensity]{\footnotesize {\it Left}: A cartoon of a muon bundle traveling through the ice.  As it reaches deeper slant depths, the muons in the bundle range out (where the arrows end).  The intensity of Cherenkov photons emitted from the bundle, as a function of perpendicular distance from the track axis, is sketched as blue curves for 
several slant depths.    
This intensity is fit to a function of the form of Eq.~\ref{eqn:kldf} with the muon bundle reconstruction \cite{Rawlins:thesis}.
{\it Right}: Lateral distribution of detected photons from an experimental event from August 2008,
together with the results of the muon bundle reconstruction's fit.}
\ec
\end{figure}

The expected signal for a given DOM can be described by an in-ice lateral 
distribution function (LDF), which depends upon its perpendicular distance, $d$,
from the center of the muon bundle, as well as the slant depth, $X$, of the muon bundle at the
point where it passes closest to the DOM, and the effective attenuation length, $\lambda_\mathit{eff}$,
of the ice surrounding the DOM.  The overall expected signal, $\xi$, is
\be
\xi = NN_{\mu, depth}(X)\frac{1}{\sqrt{\lambda_\mathit{eff}(z) d}}e^{-d/\lambda_\mathit{eff}(z)},
\label{eqn:kldf}
\ee
where $N_{\mu, depth}(X)$ is a "range-out" function which describes the number of muons as a function of slant depth:
\be
N_{\mu, depth}\left(X\right) = A\left[\left(\frac{a}{b}\right)\left(e^{bX}-1\right)\right]^{-\gamma_{\mu}}.
\ee
based on a model for continuous and stochastic energy loss of muons (parametrized
by constants $a$ and $b$ respectively), combined with the muon energy spectrum index $\gamma_{\mu}$.
As the bundle penetrates deeper, correspondingly
less light should be expected at the deeper DOMs as muons ``range out'' 
(as shown conceptually in Fig.~\ref{kath_adc_concept}):

\begin{figure}[tbp]
\begin{centering}
\includegraphics[width=.5\textwidth]{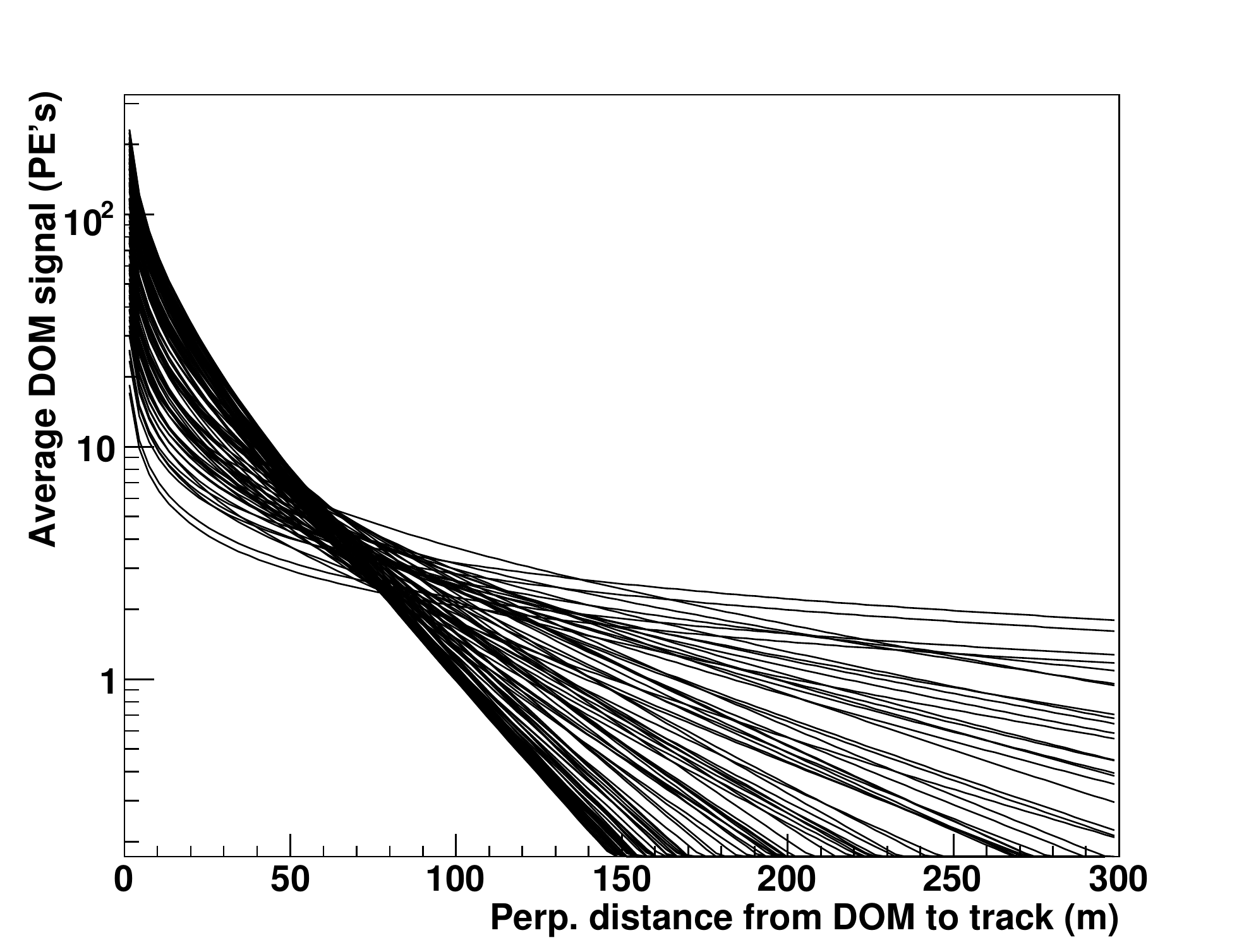}
\caption[``Bundle of Sticks'': The most stable distance for the $K$-parameter]{\footnotesize An example of a test to see the effect of possible misreconstructions.  The true track position of a single simulated 12.3 PeV shower was anchored at the surface while the direction was varied repeatedly within $1.1^{\circ}$ of the true direction.  Using a number of similar tests, the reference distance is chosen 
in order to have the smallest variation in the DOM average response: 
for this detector 70~m proved to be the most stable reference point; thus \K{} is used for the analysis.}
\label{k_sticks}
\end{centering}
\end{figure}

The effective attenuation length $\lambda_\mathit{eff}(z)$ is a combination of both scattering and absorption effects in the ice.
Both of these properties vary with DOM location, due to horizontal dust layers.
Therefore,
$\lambda_\mathit{eff}$ is a function of each DOM's depth in the ice, $z$, and is
parametrized as an average, or ``bulk,''  $\lambda_0$ times a $z$-dependent correction factor:
\be
\lambda_\mathit{eff}(z) = c_\mathrm{ice}(z) \cdot \lambda_0.
\label{eqn:cice}
\ee
The correction factor $c_\mathrm{ice}$ is based on scattering length data measured at many different depths.

For DOMs which are far from the muon bundle ($d >$~80~m),
light is likely to have traversed multiple dust layers before reaching the DOM.
Thus, in Eq.~(\ref{eqn:kldf}) the average effective attenuation length of the bulk ice, $\lambda_0$, (with no
dust correction) is used for these DOMs, and the normalization,~$N$, 
is modified by a factor chosen to ensure continuity of the function
over all distances.  

Therefore, the expected signal $\xi$ for each DOM can be computed using Eq.~(\ref{eqn:kldf}), which 
already includes the effect of dust layers and muon range-out according to that DOM's
position. 
This expected signal is then compared with the hits recorded by the DOMs using a 
likelihood method, as in Fig.~\ref{k-fit}.  
The overall normalization of the function, $N$, the average attenuation length,~$\lambda_0$, 
and the direction of the central track, are fitted as free parameters.
Although $\lambda_0$ is known to be around 25~m \cite{Lundberg:2007}, this length is fit as 
a free parameter
because minor errors in the track reconstruction can change this fitted slope of the LDF of photons
in IceCube, 
as shown in Fig.~\ref{k_sticks}, for which the direction of a single simulated 12.3~PeV track was varied from the true direction by up to 1.1$^\circ$ to show the effect of possible misreconstructions.  

After the fitted track is found, a parameter is needed to quantify the overall
amount of light (and number of muons) in the ice.  Like the \s{} parameter
(which is the IceTop lateral distribution function fit at a reference
distance of 125 m), the \K{} parameter is the IceCube lateral distribution
function of Eq.~\ref{eqn:kldf} evaluated at a reference slant depth of 1950~m,
with a reference dust correction ($c_\mathrm{ice}$ = 1),
and a reference perpendicular distance $D_{\mathrm{ref}}$ of 70 m:

\be
K(D_{\mathrm{ref}}) =\frac{A}{\sqrt{D_{\mathrm{ref}}~\lambda_0}}e^{-D_{\mathrm{ref}}/\lambda_0}
\label{eqn:k70}
\ee
where $A$ (the overall normalization) and $\lambda_0$ (the average attenuation length)
are free parameters which were fit.  
The reference distance
of 70~m was chosen to provide the most stable measurement under errors in the
reconstructed track direction.  A number of tests were performed to choose this reference point: one example is shown in Fig.~\ref{k_sticks}, where the small spread around 70~m indicates the stability of this choice in reference distance.

\subsection{Full Reconstruction: Multiple Iterations}

The IceCube track reconstruction algorithm described in the previous section 
can clearly be used to estimate the multiplicity of the muon bundle, since this
should scale with the overall normalization $A$ of the LDF of Eq.~\ref{eqn:kldf}.
But since the track direction affects the expected $\xi$'s at all the DOMs
(by changing their perpendicular distances $d$ and slant depths $X$), the same 
likelihood function can also be used to find the track direction.
The core position and direction from the surface reconstruction are used as the 
``seed track'' for the fit; the core position is anchored at the surface, 
while the direction of the track is allowed to vary in search of the best likelihood.
Because the DOMs participating in the fit are far from where the track is fixed
at the surface, a
long ``lever-arm'' gives improved angular resolution.  
Once this IceCube reconstruction has been performed, the resulting fitted track can be used as
an improved seed for a second iteration of the \emph{surface} reconstruction,
leading to a more accurate core location.  
The improved surface reconstruction is then passed back as a seed
to the IceCube reconstruction one last time.
Thus, two iterations of both the lateral fit reconstruction and the muon bundle reconstruction are performed 
(IceTop $\rightarrow$ IceCube $\rightarrow$ IceTop $\rightarrow$ IceCube). 
Figure~\ref{ICRC_res} shows the improvement to both the core location and angular reconstruction achieved through this iterative procedure.  
The two observables which will be used for composition (\K{}, which is related to the number of muons
in IceCube, and \s{}, which is related to the number of electrons and photons in IceTop) are also drawn from the
last iteration of fits, where the improved resolution allows for accuracy in measuring those parameters.  

\label{sec:multi_iter}
\begin{figure}
\begin{centering}
\subfigure[]{
\includegraphics[width=.45\textwidth, angle=0]{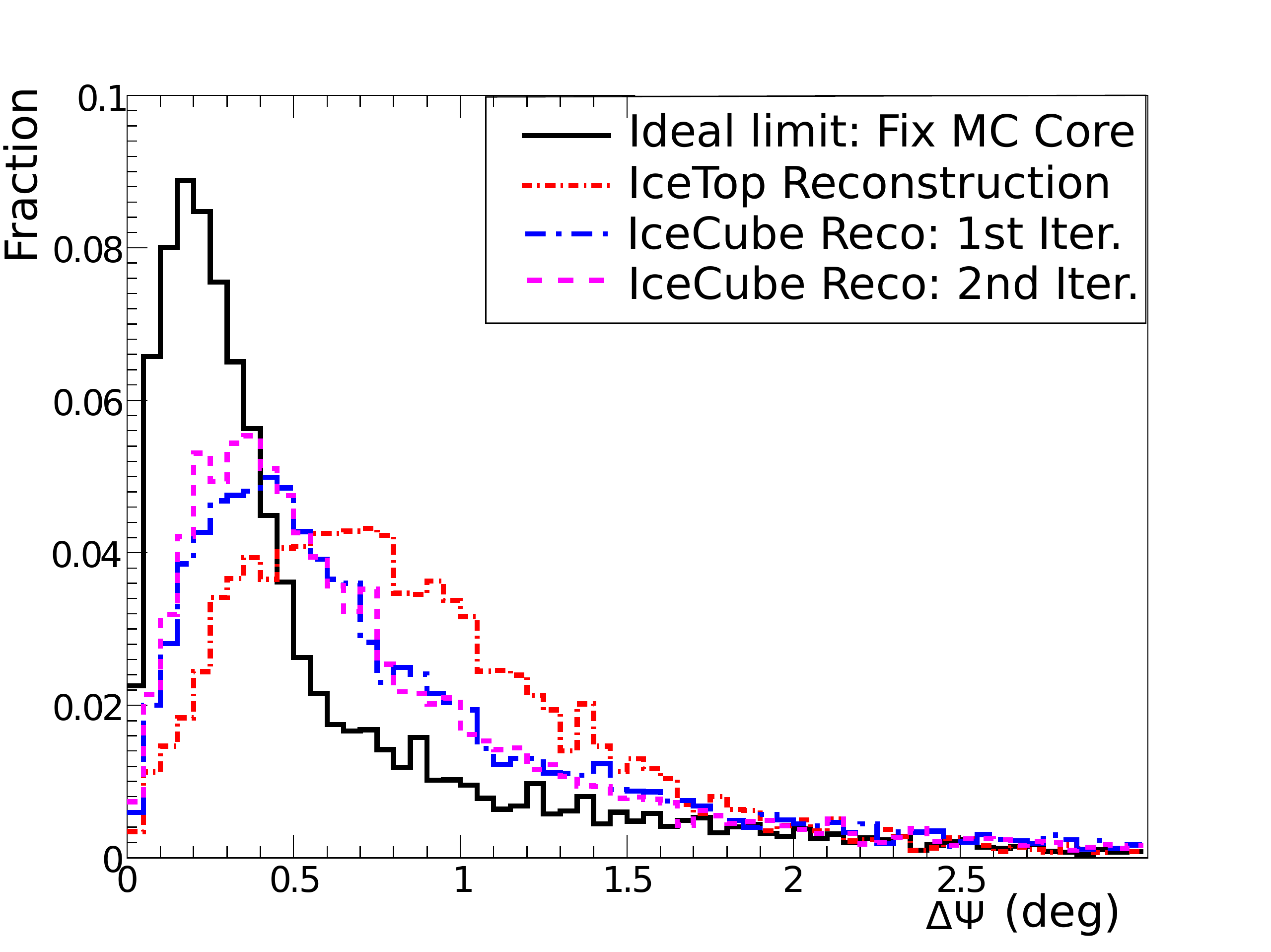}
\label{angres_icrc09}
}
\subfigure[]{
\includegraphics[width=.45\textwidth, angle=0]{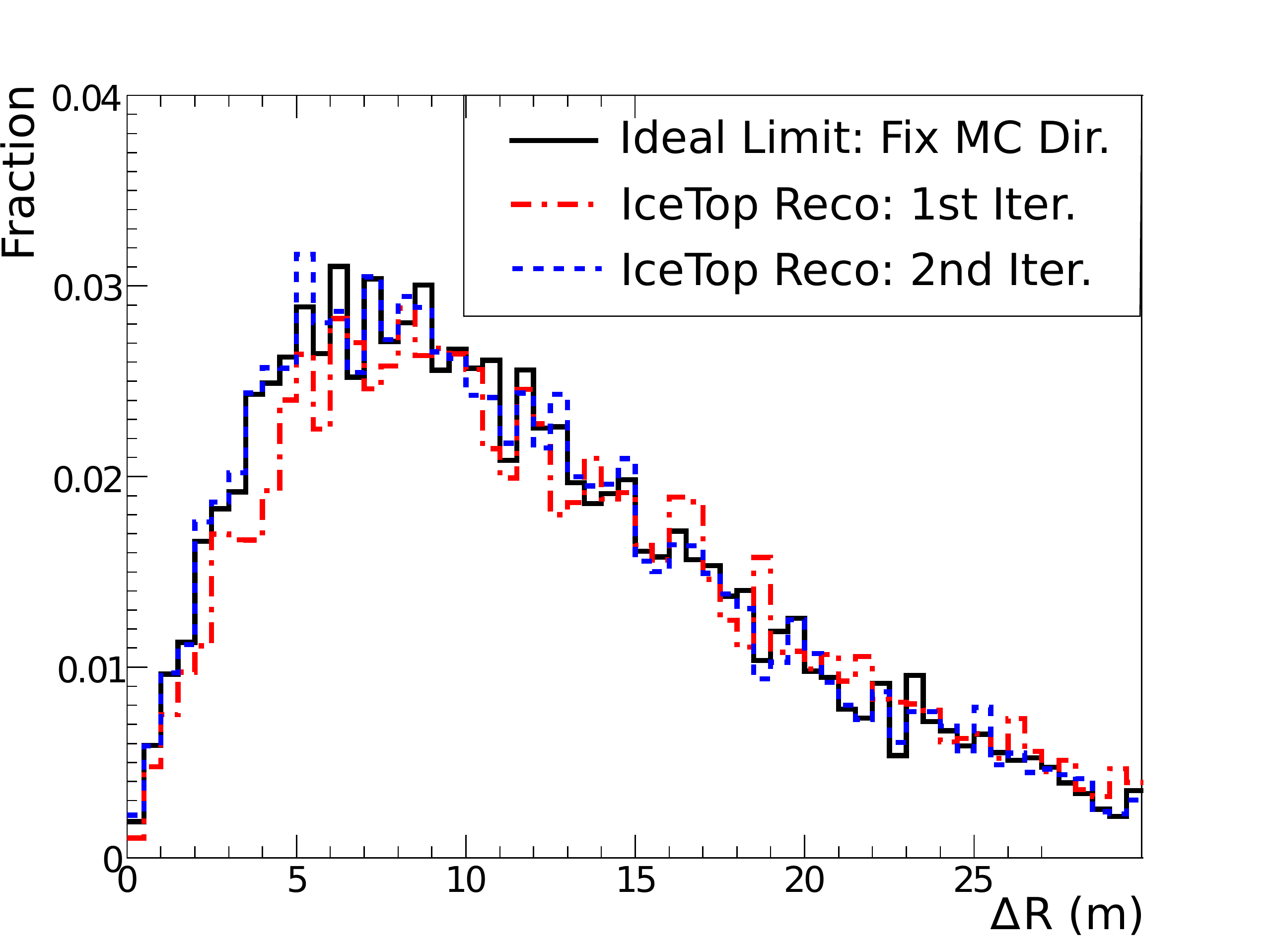}
\label{coreres_icrc09}
}
\caption[Multiple iterations]{\footnotesize  The angular resolution ({\it left}) and distance between true and reconstructed core position ({\it right}) for simulated data.  The ``ideal limit'' (black) is the maximum resolution that could be achieved by the IceCube track reconstruction ({\it left}) and by the IceTop core reconstruction ({\it right}), obtained by using the fixed true core location and true direction respectively. After two iterations of both IceTop and IceCube reconstructions, the optimal core resolution is already found, so the direction reconstruction is also limited to two iterations.
The second iteration through the two reconstructions clearly provides an improvement to both the angular reconstruction of the track and the shower core position at the surface \cite{Feusels:2009}.}
\label{ICRC_res}
\end{centering}
\end{figure}

\subsection{Snow on IceTop}
\label{sec:snow}
During the course of this analysis, it was discovered that the presence of research buildings in the vicinity of the IceTop array have caused the snow to drift unevenly over the IceTop tanks.  Some tanks are covered by more than 1~m of drifted snow, while others have a mere sprinkling: in particular, the part of the detector constructed first and nearest to structures (the ``older half'') tends to have a great deal more snow than the part constructed more recently (the ``newer half'').  Therefore, the signals in the tanks of the older half of the array are attenuated, 
and this effect is not present in the standard simulation used for this analysis.  
However, a small sample of simulated proton and iron showers \emph{with} snow attenuation 
was used to study this effect and
to develop a method of accounting for it in event reconstruction.  For each tank, the signal expectation is lowered by a factor which 
depends upon the measured snow depth above each tank, as well as the snow density, which was measured to be about 0.4~g/cm$^3$.
Thus, a correction has been made to the data at this reconstruction stage so that it corresponds to the standard simulation described above. 
The error introduced by this correction will be addressed in Section~\ref{sec:systematics}.

\section{Event Selection}
\label{sec:cuts}

\begin{figure}[tbp]
\begin{centering}
\subfigure[]{
\includegraphics[width=.45\textwidth]{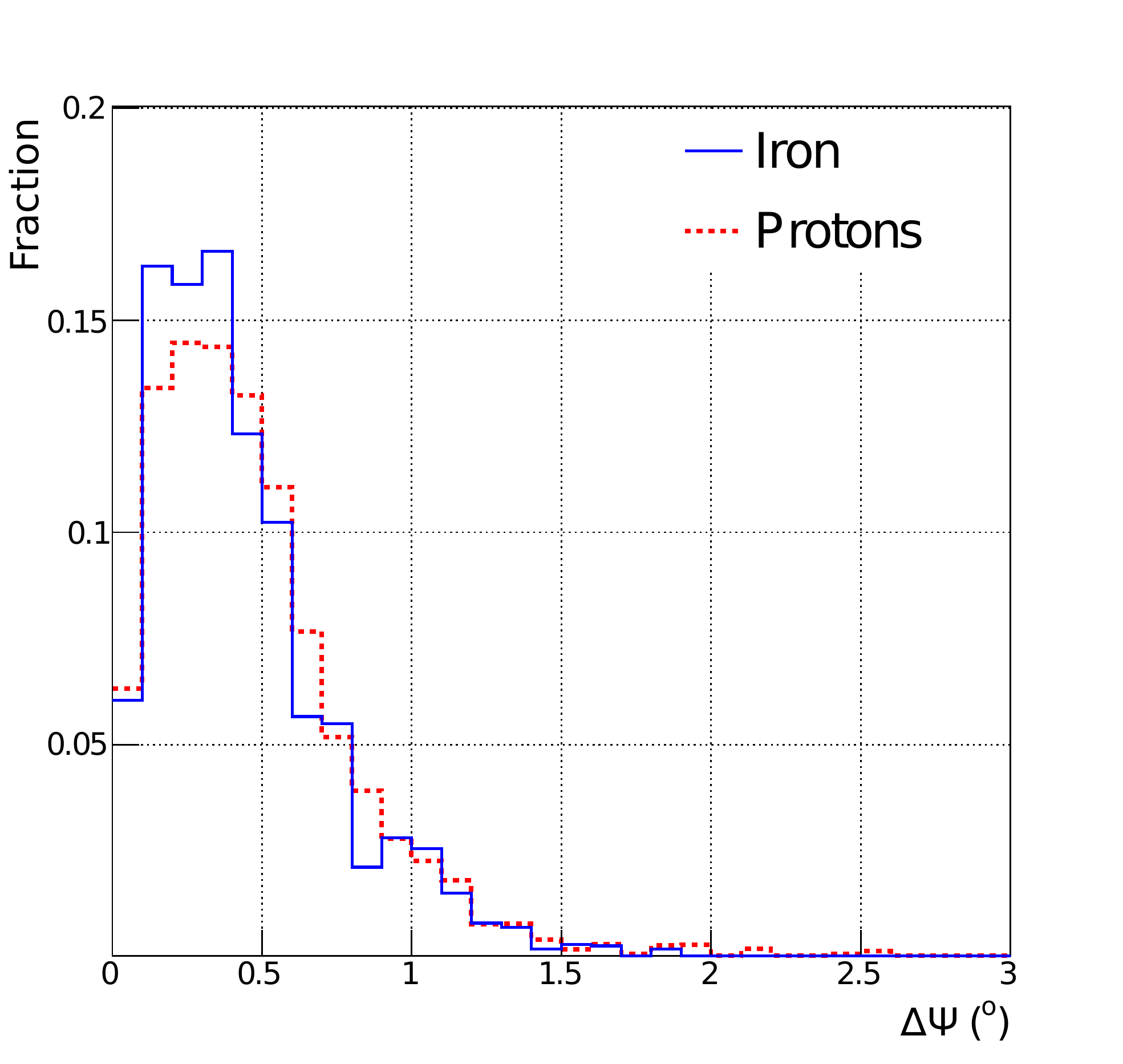}
\label{ang_diff_110pev}
}
\subfigure[]{
\includegraphics[width=.45\textwidth]{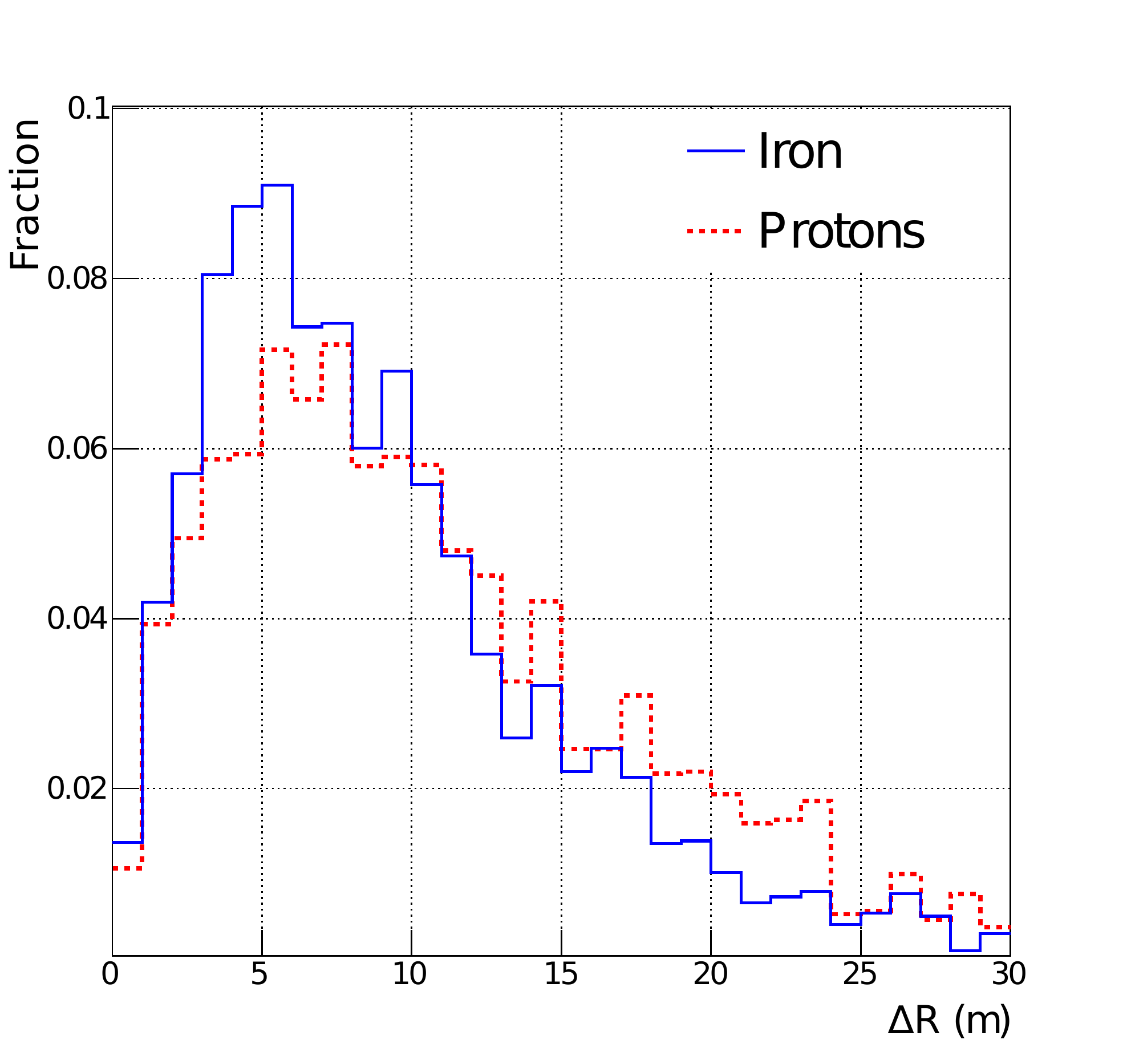}
\label{pos_diff_110pev}
}
\caption[Offsets After Cuts]{\footnotesize  The angular resolution ({\it left}) and core position resolution ({\it right}) between the true track and the final reconstructed track in proton (red dashed) and iron (blue solid) simulation after the event selection described in Section~\ref{sec:cuts}.  An angular resolution of 0.5$^{\circ}$ and a core resolution of 9~m are achieved (where resolution is defined such that 68\%, or one gaussian sigma, of all events are contained within that range).} 
\label{core_angle_cuts}
\end{centering}
\end{figure}

Experimentally observed and simulated events have been selected in multiple steps in order to ensure a reliable reconstruction of direction and core location, to retain as many events as possible, and to avoid introduction of any bias by primary mass and energy:

\begin{itemize*}
\item {\it Step 1: Initial filter and reconstruction requirements.}  For this analysis, a simple filter was applied requiring 5 IceTop stations and 5 IceCube DOMs for an event to pass on to the next stage.  (Note that although 8 IceCube DOMs were required to trigger IceCube, some of those hits will have been removed in a cleaning phase where hits are clustered into events.  This process can leave less than 5 DOMs in the final event.)  Furthermore, for some events either the IceTop or IceCube reconstruction (as discussed in Section~\ref{sec:multi_iter}) fails to find a successful minimum.  At this stage, an event is excluded if it does not have successful reconstructions in both IceTop and IceCube.

\item {\it Step 2: The shower and the track are reconstructed as passing within the area and volume of the two arrays.}  For an event to be considered ``contained'' within IceTop, the shower core must be reconstructed within the two-dimensional outer boundary of the array.  To satisfy IceCube containment, the track must be reconstructed within the three-dimensional outer boundary of IceCube.  This selection is applied using three different reconstruction algorithms: the final muon bundle reconstruction (which is used for the analysis), the first surface reconstruction (which has not been biased by any IceCube track reconstructions), and an independent IceCube track reconstruction (which has not been biased by taking as input a surface reconstruction as a first guess) must all satisfy IceTop and IceCube containment.  As shown in Table~\ref{table:cut_stats}, this basic selection eliminates the largest fraction of events, but is the most important one for improving core reconstruction and angular resolution.

\item {\it Step 3: The final IceTop reconstruction and the final IceCube reconstruction show good directional agreement.}  A disagreement as to the shower direction between the final iteration of the surface and IceCube track reconstructions indicates a shower for which either \s{} or \K{} was reconstructed poorly and it is impossible to discern which, if either, is correct.  The opening angle, $\Delta\Psi$, between the surface reconstruction and the IceCube track reconstruction 
is conservatively required to be less than 1$^{\circ}$.

\item {\it Step 4: The length of the track within the IceCube array must be reasonably long.}  High energy showers with small track lengths are likely to be events with a poorly reconstructed direction and/or core position.  Thus, a requirement that a track length -- estimated by calculating the length of the track segment between the first and last direct hit if within a -15~ns to +75~ns residual time window -- is greater than 850~m for showers with \s>5.625, linearly decreasing to 400~m at \s{} = 0, is applied.  

\item {\it Step 5: The fitted slope of the LDF function is within a reasonable range.}  The reconstructed effective propagation length, $\lambda_{0}$, should resemble the independent measurements for Antarctic ice, though a range of values can still provide a robust \K, as described in Section~\ref{sec:K70} and shown in Fig.~\ref{k_sticks}.  As this parameter is a free parameter in the fit, it does have a slight dependence on the mass of the primary particle; thus, to avoid introducing a bias into the event sample very loose requirements are made:
\be
\label{Eqn:Length}
20~\mathrm{m} < \lambda_{0} < 150~\mathrm{m}~~;~~\lambda_{0} < \frac{100}{K_{70}} + 40~~;~~\lambda_{0} < \frac{150}{S_{125}} + 40. 
\ee

\item {\it Step 6: Events containing random coincidences are removed.} 
In experimental data, uncorrelated events which pass either the IceTop or the IceCube detector can mimic a coincident air shower.  Such events are not simulated.  These events can be isolated by examining a parameter for the arrival time difference of the track hypothesis at the surface of the ice:
\be
\Delta t = \frac{z_{IT} - z_{IC}}{c \cdot \cos(\theta_{IC})} - (t_{IC} - t_{IT}),
\ee
where $z_{IT}$ is the altitude of the reconstructed core position of the shower in IceTop, $z_{IC}$ is the point on the IceCube track closest to the center of gravity of the hits, $t$ is the reconstructed arrival time of the shower  at depth $z$ and $\theta$ is the zenith angle, as reconstructed by independent IceCube (IC) and IceTop (IT) reconstructions.  It has been found that these uncorrelated events can be distinguished from the truly coincident sample by requiring $\Delta t$ less than 3~$\mu$s.
\end{itemize*}

\begin{table}[tbp]
\begin{center}
\begin{tabular}{cccccc}
\toprule
Event	&	\multicolumn{5}{c}{Number of Remaining Events}\\
\cmidrule{2-6}
Selection	&	Data	&	Simulation	&	 Contained	&	$\Delta\Psi$ < $1^\circ$ 	&	$\Delta$R < 25~m	\\
\midrule
Step 1	&	1141590	&	62954	&	16601	(26.4~\%)	&	25490	(40.5~\%)	&	35313	(56.1~\%)\\
Step 2	&	368815	&	11737	&	11201	(95.4~\%)	&	10443	(89.0~\%)	& 	11190	(95.3~\%)\\
Step 3	&	325685	&	10959	&	10484	(95.7~\%)	&	~9993 	(91.2~\%)	& 	10532	(96.1~\%)\\
Step 4	&	259717	&	8392	&	~8129	(96.9~\%)	&	~7872	(93.8~\%)	&	~8074	(96.2~\%)\\
Step 5	&	239902	&	8122	&	~7868	(96.9~\%)	&	~7694	(94.7~\%)	&	~7815	(96.2~\%)\\
Step 6	&	239797	&	8112	&	~7859	(96.9~\%)	&	~7685	(94.7~\%)	&	~7808	(96.3~\%)\\
\bottomrule
\end{tabular}
\end{center}
\caption[]{\footnotesize Event passing rate at each selection step (see Section~\ref{sec:cuts}).  In this table, simulation includes proton and iron only, of which 6.6 million total events were generated.  The event selection refers to the steps in Section~\ref{sec:cuts}. ``Contained'' shows the number of simulated events at each step for which the true simulated track was contained within the area and volume of the two arrays.  $\Delta\psi$ is the opening angle between the true and the reconstructed direction, and $\Delta$R is the difference between the true and the reconstructed core location.  In each case, at each cut level, the number of events is given as well as the percentage they comprise of the total number of events at that cut level.  After all six selection levels, at least 94\% of the events used for this analysis are well-contained and have a core resolution within 25~m and an angular resolution within $1^{\circ}$. 
Simulation events listed here are not re-weighted from their $E^{-1}$ generation spectrum. }  
\label{table:cut_stats}
\end{table}%

With this event selection, the resolution in core position is better than 9~m and in track direction is less than 0.5$^{\circ}$ (where resolution is defined such that 68.3\% of all events are contained within that range), as shown in Fig.~\ref{core_angle_cuts}.  The final event sample contains 239797 events from the August 2008 experimental data and 20289 for all five simulated primaries from the Monte Carlo simulation.

\begin{figure}[tbp]
\begin{centering}
\includegraphics[width=0.5\textwidth]{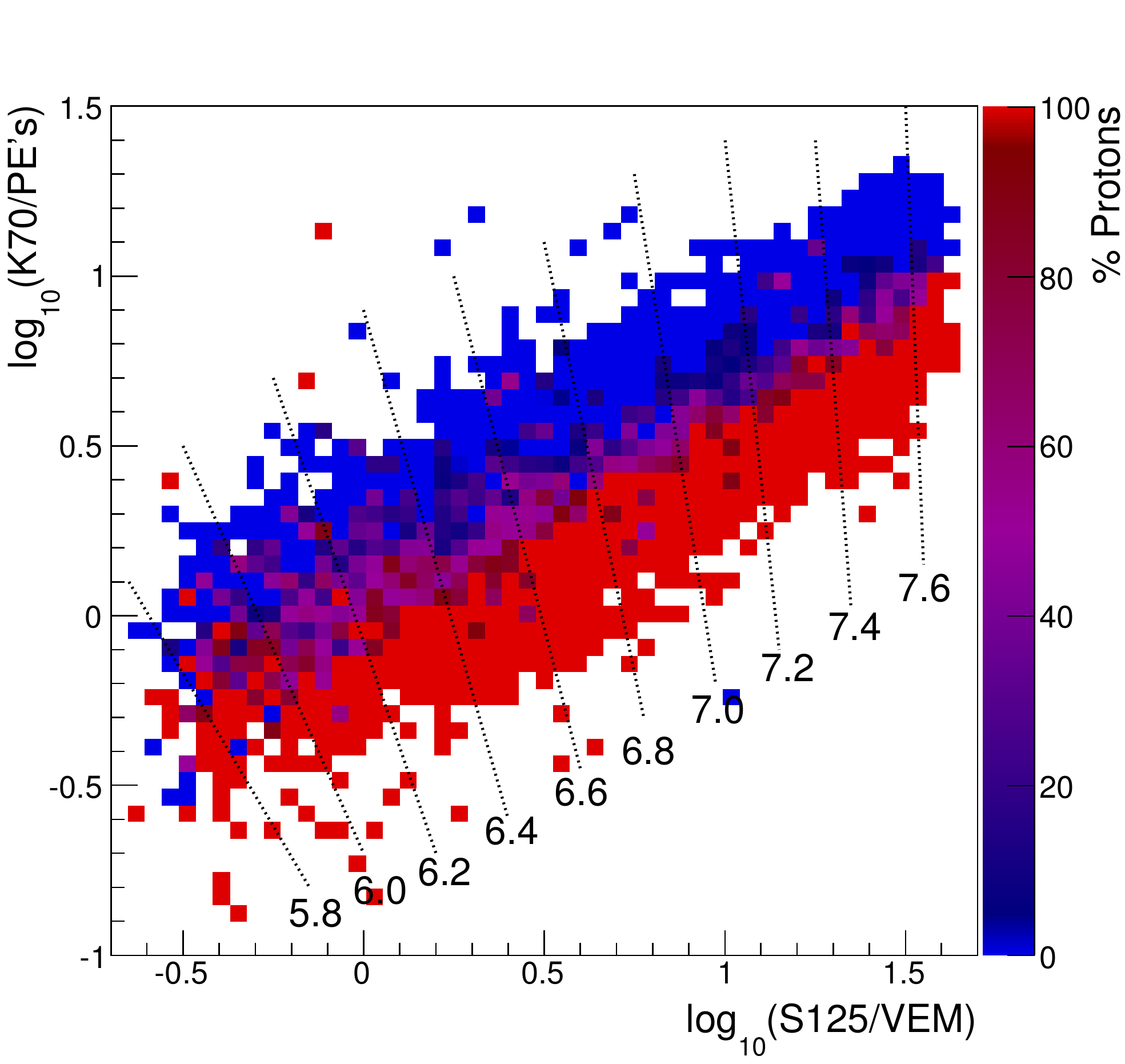}
\caption[Data and Simulation in \K{}-\s{} Space, Showing the Percentage
of Proton and Contours of True Energy (the ``pig plot'')]{\footnotesize 
The \K{}-\s{} parameter space after the final event selection.  The colors depict the percentage of protons in each bin (where for clarity only simulated proton and iron showers are shown).  A bin which has more proton than iron will appear reddish, 
whereas a bin containing more iron than proton will be shaded in blue.  
Intermediate purples indicate regions of overlap.
The dotted black lines depict approximate contours of constant primary energy and are labeled in log$_{10}$(E$_{0}$).    (For the analysis, all five primary types were included)
}
\label{berries_eps}
\end{centering}
\end{figure}

\section{Neural Network Mapping Technique}
\label{sec:nn}

\begin{figure}[tbp]
\begin{centering}
\includegraphics[width=.5\textwidth]{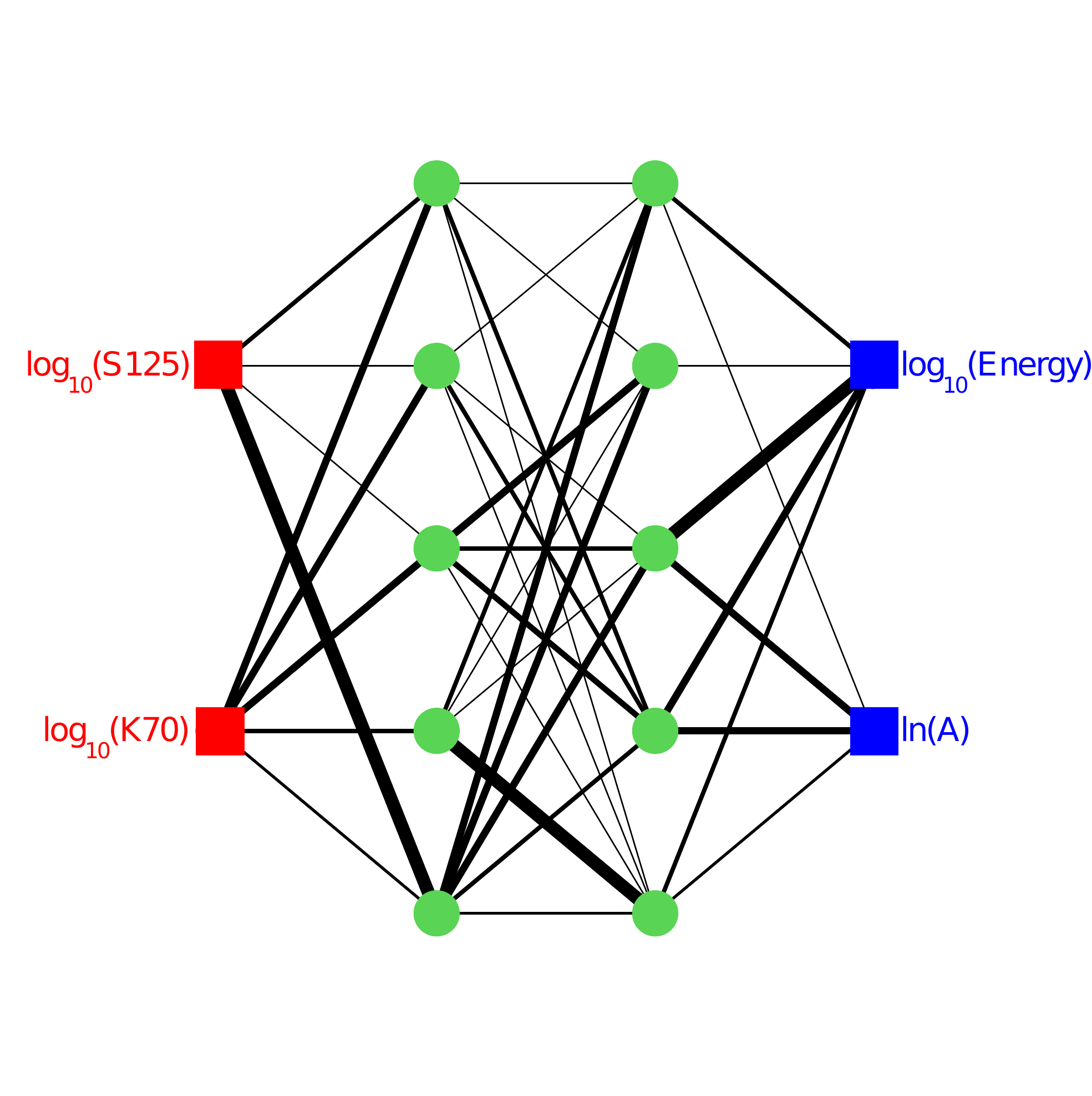}
\caption[Neural Network Structure and Error Function]{\footnotesize
This figure shows the neural network chosen, with structure 2:5:5:2.  The nodes are depicted with (green) circles, and the black lines are the weights connecting the nodes.  Weights are determined by training the network; thicker lines represent a stronger connection between nodes.  
}
\label{nn_structure}
\end{centering}
\end{figure}

After the event selection criteria from Section~\ref{sec:cuts} have been applied, a strong relationship between muon and electron observables 
(the vertical axis \K, and the horizontal axis \s), and cosmic ray primary mass and energy ($A$ and $E_0$) can be seen in Fig.~\ref{berries_eps}.
Black lines approximating contours of energy (in $\log_{10}(E_0/\mathrm{GeV})$) guide the eye.
Mass is indicated by color; each colored 
bin in the \K{}-\s{} space is shaded according to the percentage of proton found in that bin for simulated proton and iron primaries only for clarity.  In the analysis intermediate species were included.  The separation between proton and iron is clearly visible in Fig.~\ref{berries_eps}, especially at the highest energies.  However, the energy contour lines are not parallel and the proton and iron distributions are also not parallel, which implies that the relation between the \K{}-\s{} space and the mass-energy space is non-linear, as the proton/iron distribution becomes distinctly more orthogonal to \s{} at higher energies. 

In light of the non-linear correlation between the first two-dimensional space (\K{}-\s) and the second ($A$-$E_0$), a mapping technique is required to step between them. 
For flexibility in testing combinations of multiple new input parameters under development in IceCube and IceTop, a neural network technique was developed.  It is important to emphasize that both outputs are continuous in nature (as opposed to quantized); therefore, this style of neural network is referred to as either an interpolation or a regression problem  \cite{Bishop:1994}.   

A neural network consists of a set of input parameters which are connected to a set of output parameters through a series of nodes which are arranged in layers.  Each node in a layer is connected to all nodes of the previous and subsequent layer via a series of weights, and is assigned an activation function which, for an interpolation problem, is typically a sigmoid function for the internal layers and a linear function for the output layer.  Each node then represents the activation function acting on the summed, weighted input parameters.\fixed{So, in the most basic terms, the outputs from a network with a single internal layer are a function of a function of an input parameter.} The structure used for this analysis can be found in Fig.~\ref{nn_structure}.

The neural network used here, shown in Fig.~\ref{nn_structure}, is a feed-forward multilayer-perceptron as found in the \textsc{ROOT~TMultilayerPerceptron} package \cite{ROOT}.  The network takes log$_{10}$(\s) and log$_{10}$(\K)~as input parameters and is trained on Monte Carlo simulation data
to find output parameters of log$_{10}$($E_0$) and ln($A$) \cite{Andeen:thesis}.
Both the two inputs and the two outputs are renormalized so that they are 
numbers between 0 and 1.

Over the course of a number of training cycles, the network ``learns'' to find the best fit to the expected outputs.  During each cycle, the error between the output parameters (obtained from the given network structure) and the true parameters is calculated.  The weights in the network are updated, as prescribed by a learning algorithm, to reduce this error.  The error is minimized after a sufficient number of learning cycles.  This iterative training procedure is performed using 1/4 of the final Monte Carlo sample ($\sim$1000 events of each species).  This ``training'' sample provides the network with \K{} and \s{} as well as the true mass and energy outputs.

A ``testing'' sample, consisting of an independent set of 1/4 of the final Monte Carlo samples, is used to check the progress of the neural network training. 
After each learning cycle, the testing sample is also passed through the network, the output parameters are calculated and then compared with the true values, to verify that the network is not becoming too specific to the training sample (in the case of ``overtraining" \cite{Bishop:1994}).
The testing set is also used to optimize a number of other network parameters, such as learning method, architecture, and activation functions.  For this analysis a broad range of combinations were evaluated and the network which  was chosen provided the testing sample with the best energy resolution and the best mass reconstruction for the energy range accessible (1~PeV-- 30~PeV), within the fewest learning cycles.  

The final optimized network had the following characteristics:
\begin{itemize}
\item {\it Architecture:} One method for choosing network architecture is to incorporate the fewest hidden layers and nodes possible to provide results of the desired accuracy \cite{Bishop:1994}.  Following this prescription, the network chosen had two hidden layers of five nodes each (for the 2:5:5:2 setup as seen in Fig.~\ref{nn_structure}).
\item {\it Activation Functions:} The architecture is defined by nodes with sigmoid activation functions for the hidden layers and linear activation functions for the output layer.  
\item {\it Learning Algorithm:} The quasi-Newtonian learning algorithm of Broyden, Fletcher, Goldfarb and Shanno (known as BFGS) was chosen based on its efficiency in returning a desirable network \cite{ROOT}.
\item {\it Training Cycles:} Using 947 training cycles provided the smallest errors without overtraining for this particular structure, learning algorithm, and simulated data sample.  
\end{itemize}

To avoid biases, the training dataset was not included in the simulation data sample used for the analysis.  
Furthermore, while the testing sample was not used to train the network, it was used to optimize the choice of the network structure, learning algorithm and the number of learning cycles, and it was used to optimize the minimization process (as will be discussed in Section~\ref{sec:comp}), all of which could introduce a bias.  Thus, the testing sample was also removed from the final sample.  
Half of the simulated data was independent of the training and testing samples ($\sim$2000 events of each species) and is called the ``analysis'' sample.  This half was used for the final analysis steps described below.  

The normalized energy output from the neural network can be translated directly back into units of $\log_{10}(E/\mathrm{GeV})$.
The energy resolution of this measurement ranges from $18-20\%$ in the threshold region ($\sim$1.5~PeV) to $6-8\%$ at 30~PeV, for an average resolution better than $14\%$ over the full range of energies, with a bias which is well within the resolution, as seen in Fig.~\ref{std_energy_plots}.  
In the analysis of energy spectrum and composition described in the following sections, events are placed in energy bins which are wider than this
energy resolution. 
The normalized mass output from the neural network, by contrast, exhibits much more spread in mass-space and is not well-resolved,
as seen in Fig.~\ref{mini_step0_5type}.  The most likely average composition of the population of cosmic rays can still be measured,
with a further minimization step to be described in Section~\ref{sec:comp}.

\begin{figure}[tbp]
\begin{centering}
\subfigure[]{
\includegraphics[width=.45\textwidth]{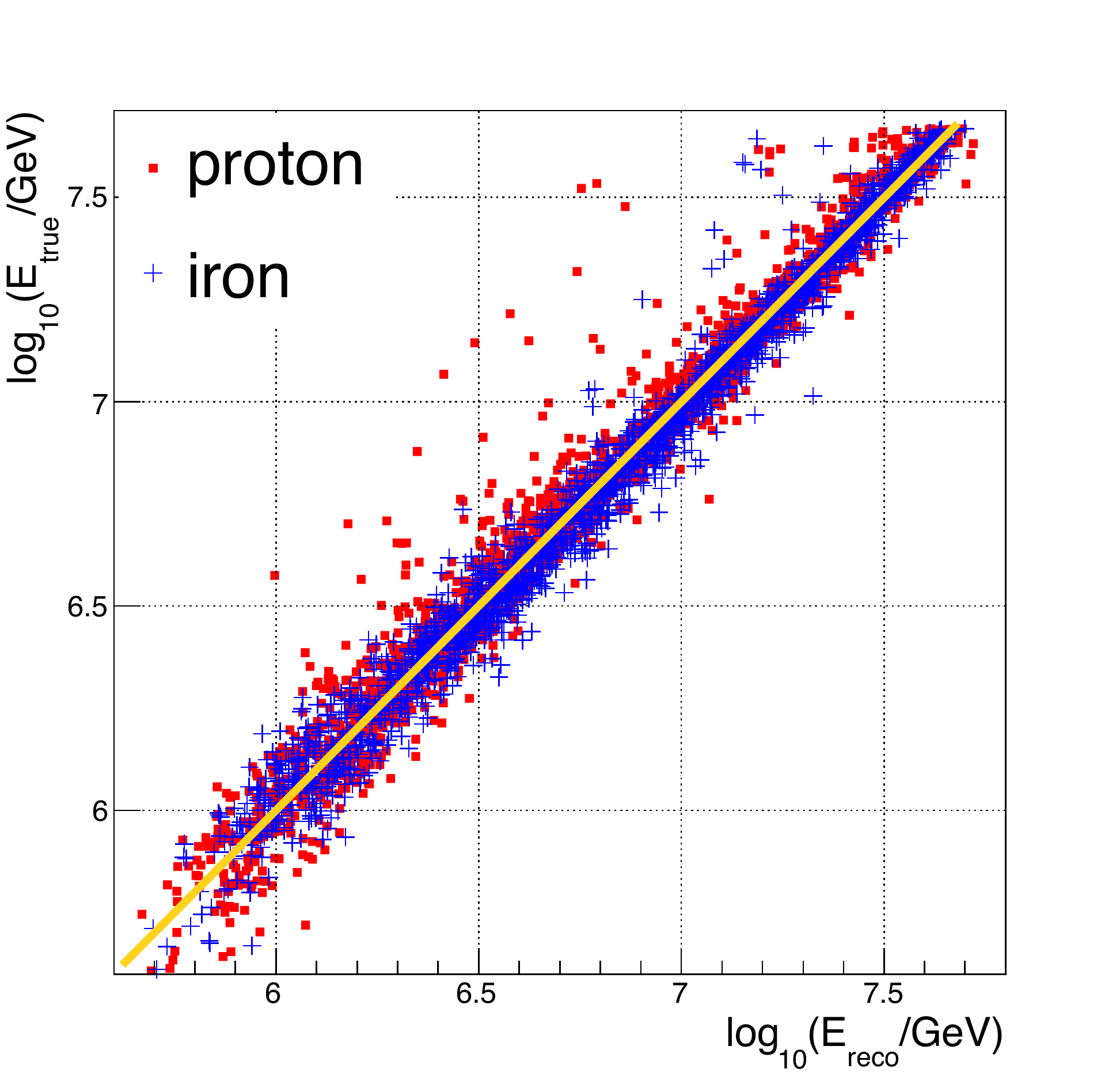}
\label{etrue_nnenergy}
}
\subfigure[]{
\includegraphics[width=.45\textwidth]{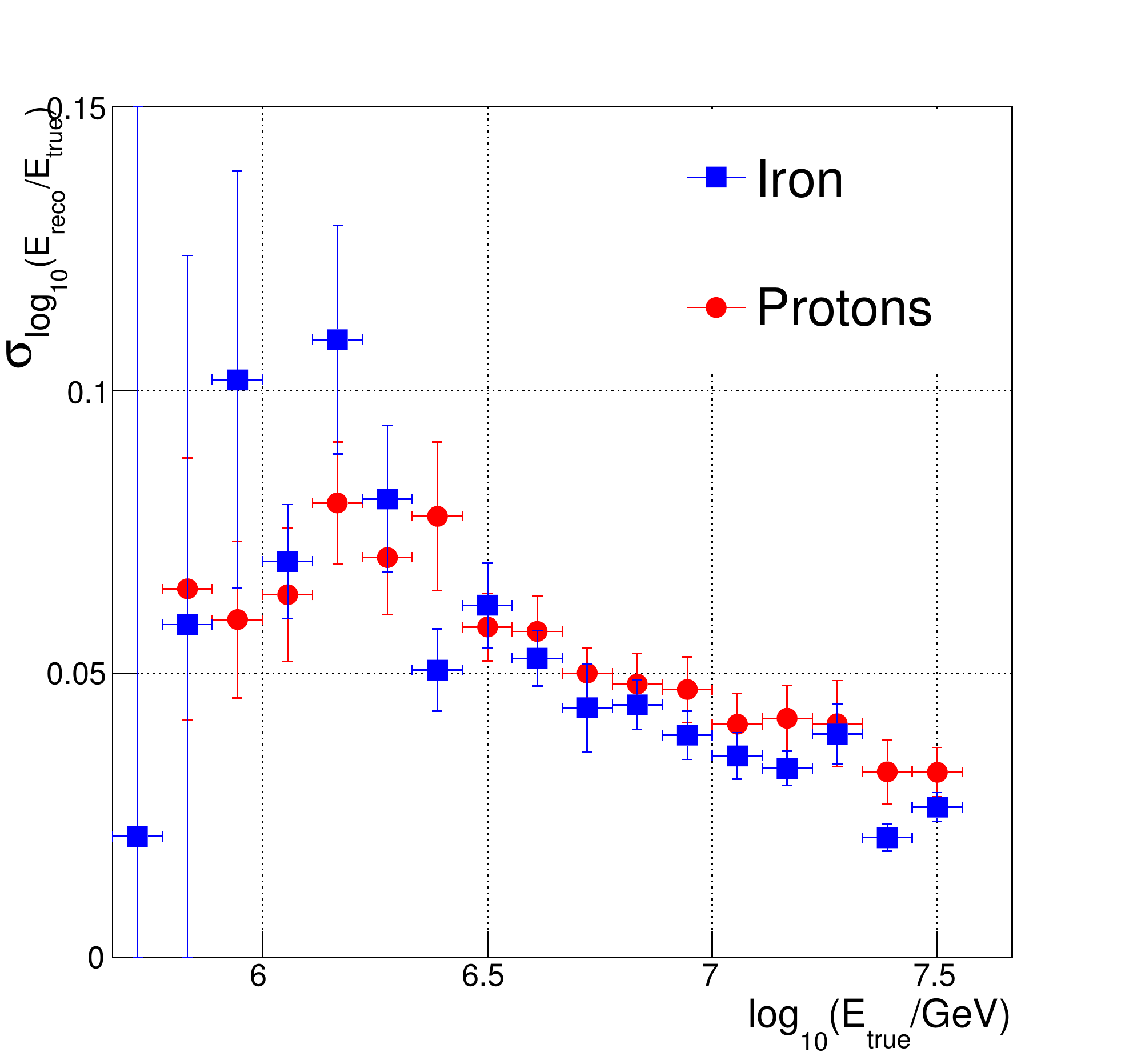}
\label{eres_vs_e}
}
\subfigure[]{
\includegraphics[width=.45\textwidth]{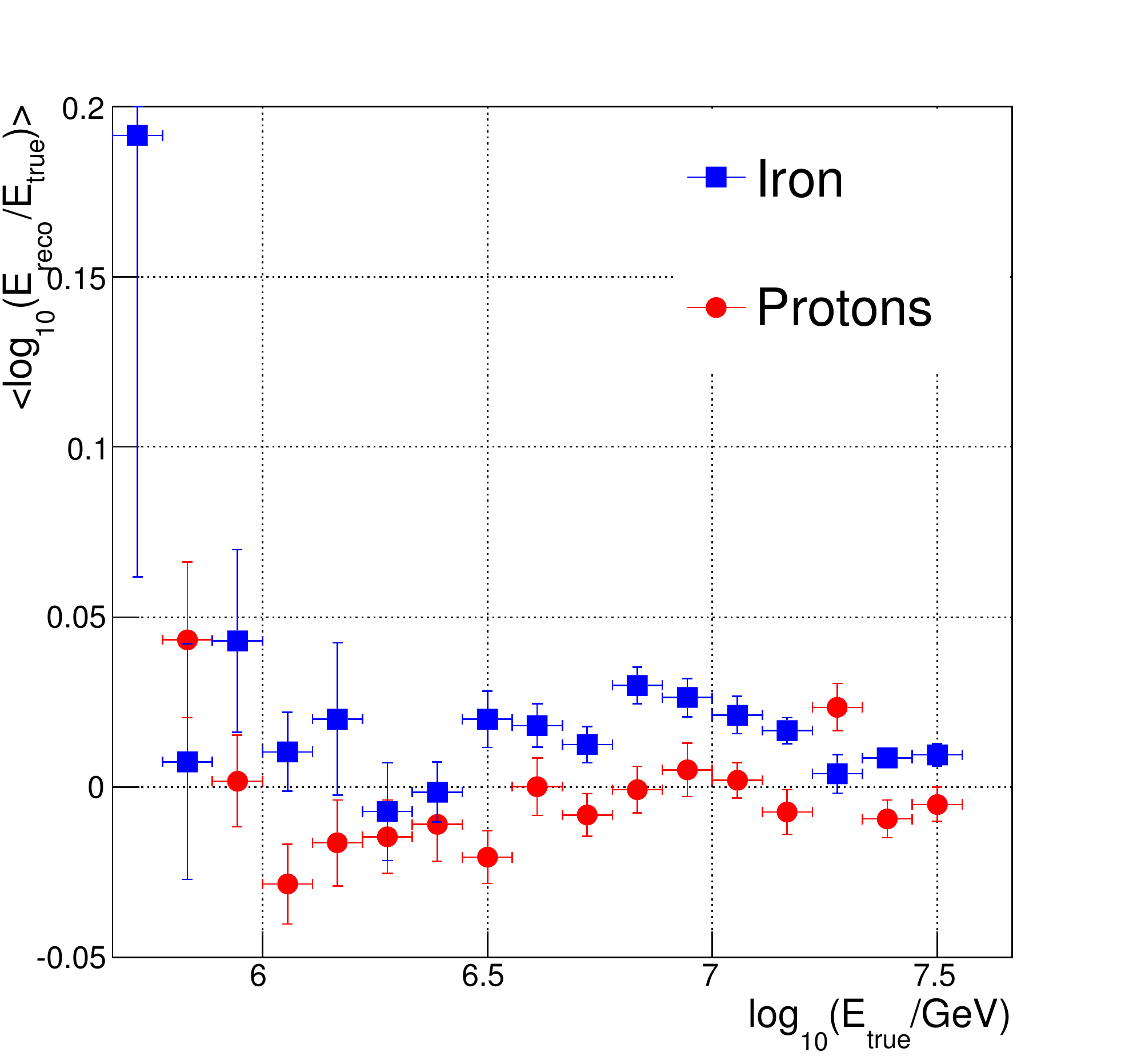}
\label{ebias_vs_e}
}
\subfigure[]{
\includegraphics[width=.45\textwidth]{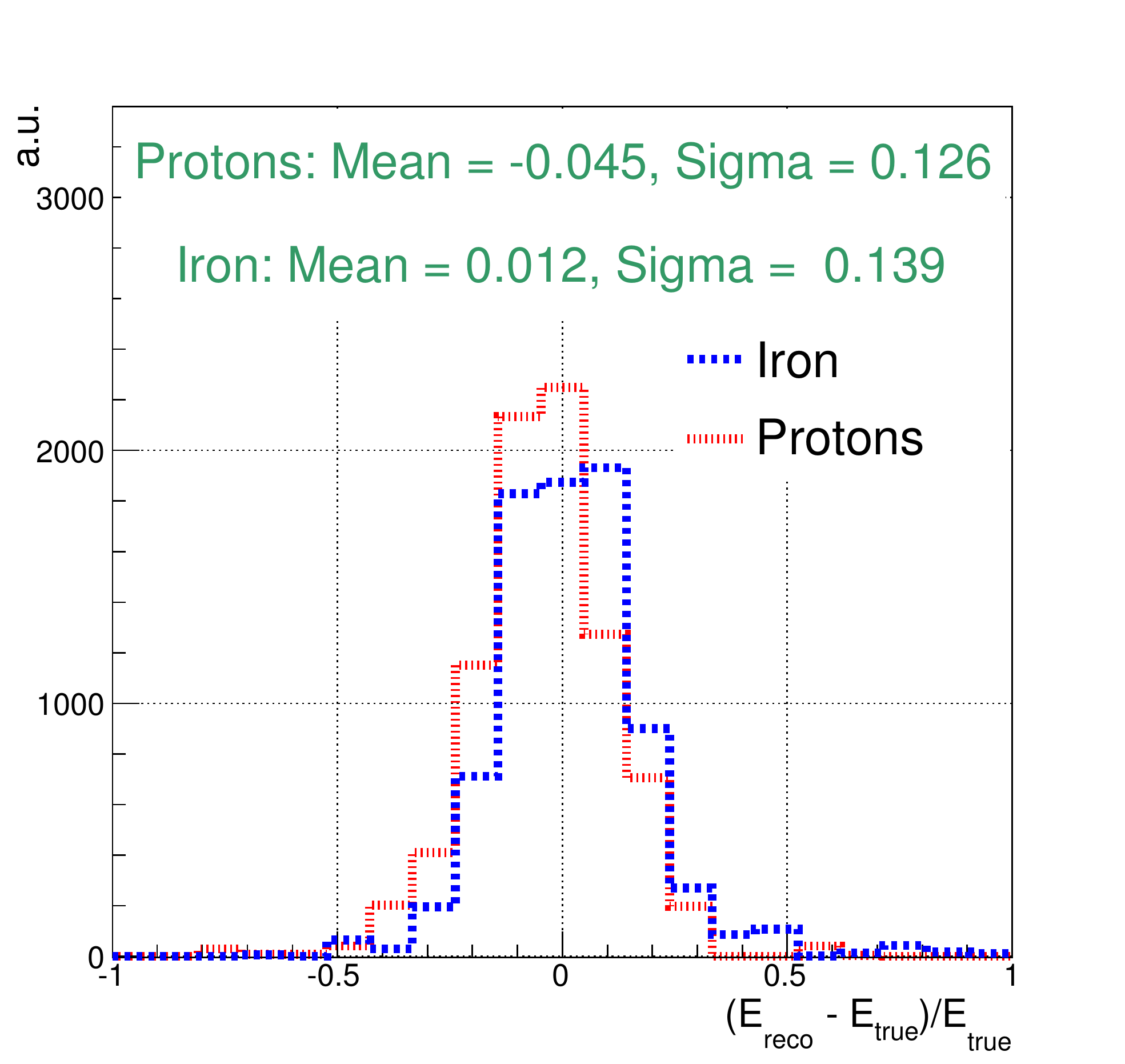}
\label{eres}
}\caption[Energy resolution and bias for the neural network output energy]{\footnotesize Performance of energy reconstruction.  In these figures using simulated data, E$_\mathrm{reco}$ is the reconstructed energy provided by the neural network, while E$_\mathrm{true}$ is the true primary energy generated for a given air shower.  Red (circles) mark proton and blue (squares) denote iron.  (a): The true vs the reconstructed energy  of the IceTop/IceCube 40-string/40-station detector, including a yellow line representing a 1-to-1 ratio to guide the eye.  (b): The energy resolution (68 percentile deviation from the 1-to-1 line), which ranges from $18-20\%$ in the threshold region ($\sim$1.5~PeV) to $6-8\%$ at 30~PeV.  (c): The energy bias or energy misreconstruction (the shift of the peak distributions from the same 1-to-1 line).  (d): The energy resolution for the full energy range used in this analysis, for both proton and iron, where the mean and sigma reported are from fits of a Gaussian distribution to each curve individually.  }
\label{std_energy_plots}
\end{centering}
\end{figure}

\section{Energy Spectrum}
\label{sec:espec}

At this point it is important to note that the correction described in Section~\ref{sec:snow} accounts for the signal attenuation due to the snow, but cannot account for the change in the composition-dependent energy threshold -- which is higher in energy for the older half of the array due to the deep snow.  Therefore, from this point forward, in both simulated and experimental data events with a shower core located on the older side of the array (x-position more than 200~m in Fig.~\ref{Tom_event_display}) have been removed to minimize this effect.  The number of events surviving as a function of reconstructed energy can be found in Table~\ref{table:nevents_finaleslices}.  

For a determination of the energy spectrum, the data were divided into energy bins.  Nine bins for each decade in energy 
are chosen so that the bin width is sufficiently larger than the resolution and bias in the reconstructed energy.

The flux, $\Psi$, as a function of energy, $E_0$, is given by: 
\be
\Psi(E_0) = \frac{1}{{\eta}A{\Omega}\tau} \frac{\mbox{log$_{10}$}(e)}{E_0} \frac{dN}{d\mbox{log$_{10}$}(E_0)}, 
\label{flux}
\ee
where $e$ is Euler's constant (the base of the natural logarithm);  
$\eta(E_0)$ is the efficiency (Fig.~\ref{eff_vs_e}), defined as the ratio of simulated events left after the final event selection to the number generated; $A=\pi$(1200~m)$^2$ is the geometrical area over which the CORSIKA air showers were generated; $\tau=$ 29.78 days is the exposure time, or livetime, of the detector; and $\Omega$ is the solid angle weighted by the generated zenith distribution:
\be 
\Omega = 2{\pi}\int_{0}^{{\theta}_{max}}\cos(\theta)\sin(\theta)d\theta = \pi \sin^{2}(\theta_{max}),\ee 
where $\theta$ is the zenith angle.  In other words, the flux can be found by dividing the energy distribution by the detector aperture, $\mathcal{A}(E_0) = {\eta}(E_0)A{\Omega}$, and the exposure time, $\tau$. 

\begin{figure}[tbp]
\begin{centering}
\subfigure[]{
\includegraphics[width=.46\textwidth]{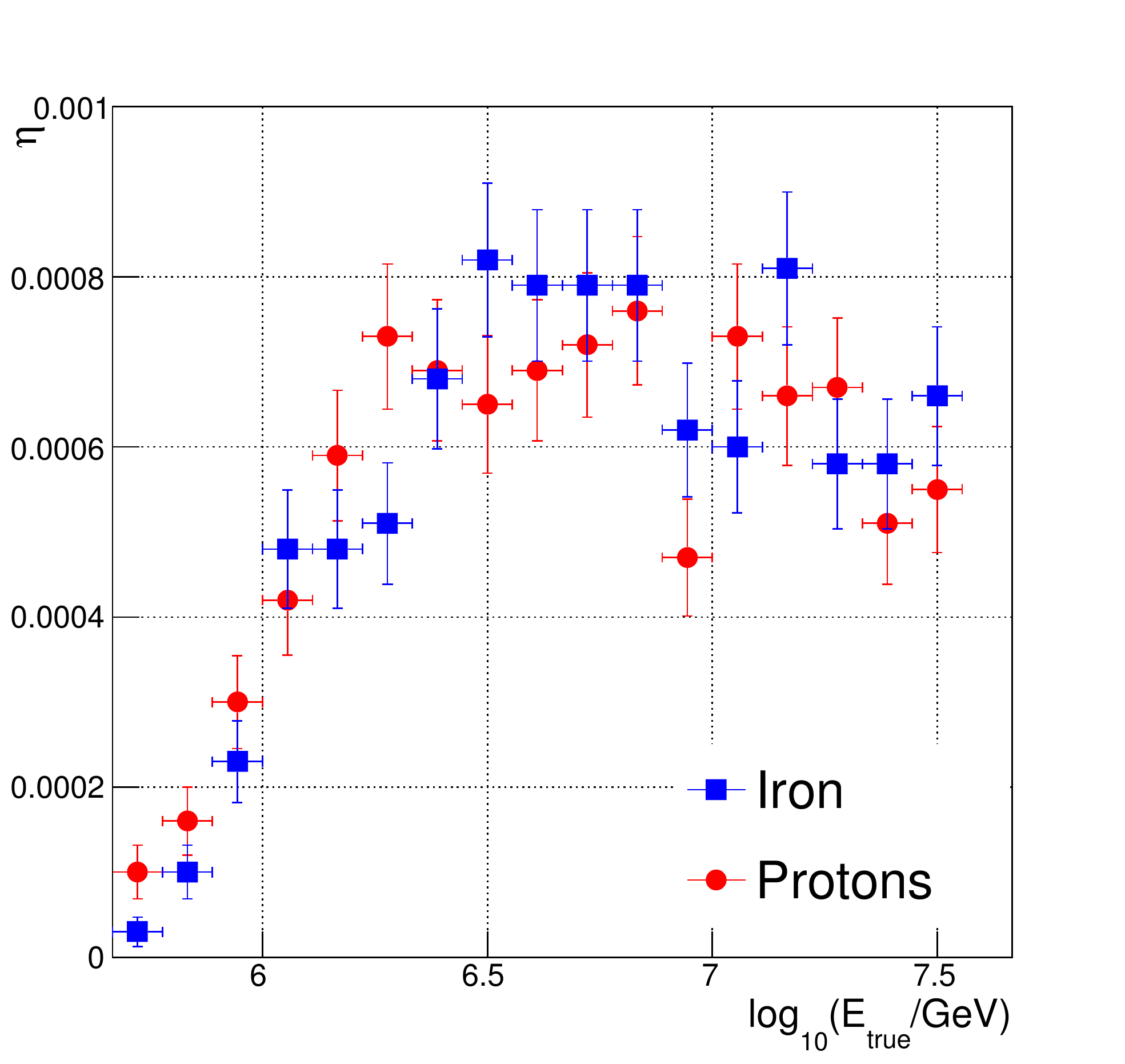}
\label{eff_vs_e}
}
\subfigure[]{
\includegraphics[height=.44\textwidth]{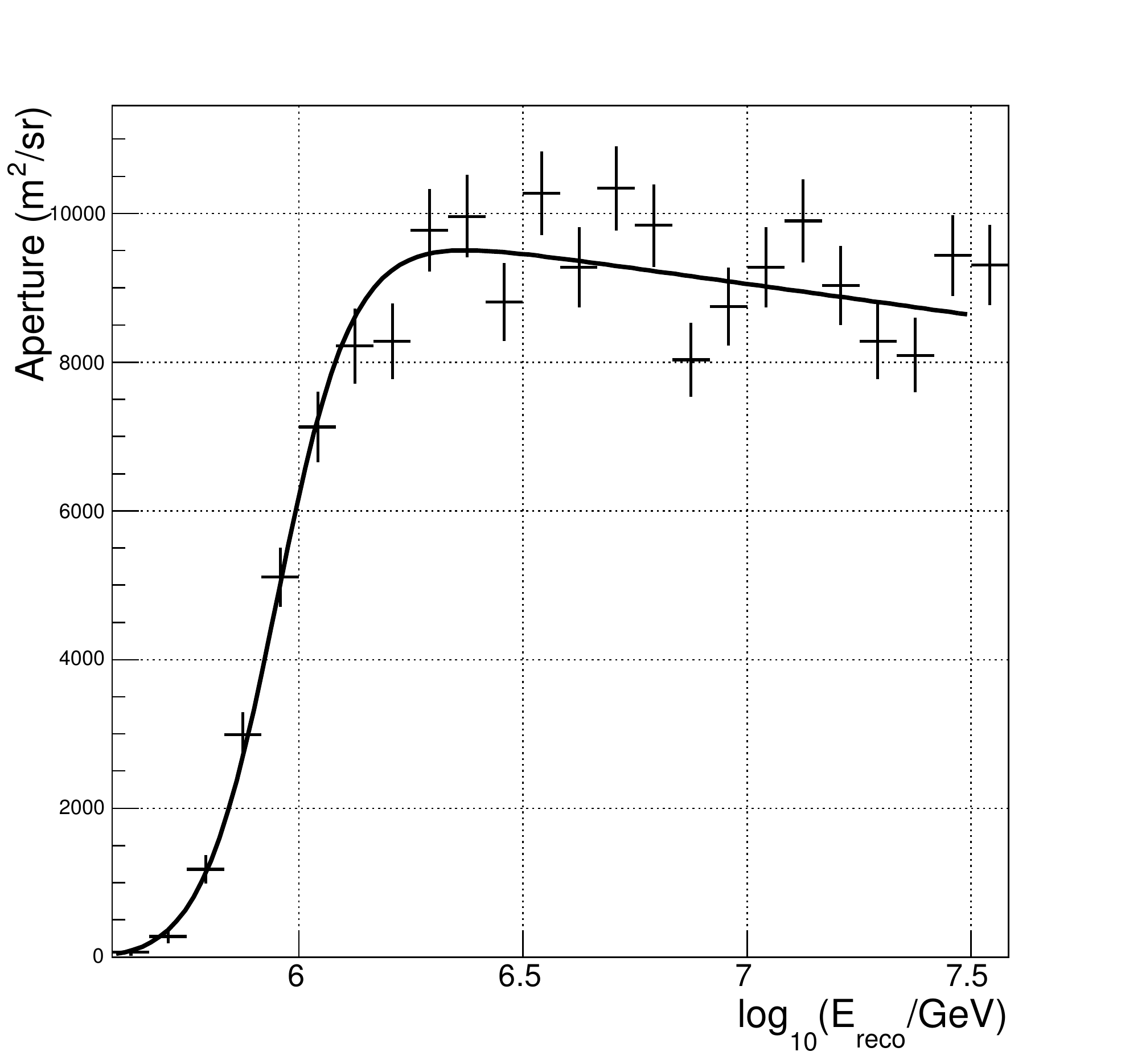}
\label{ap_reco_slanty}
}
\subfigure[]{
\includegraphics[width=.55\textwidth]{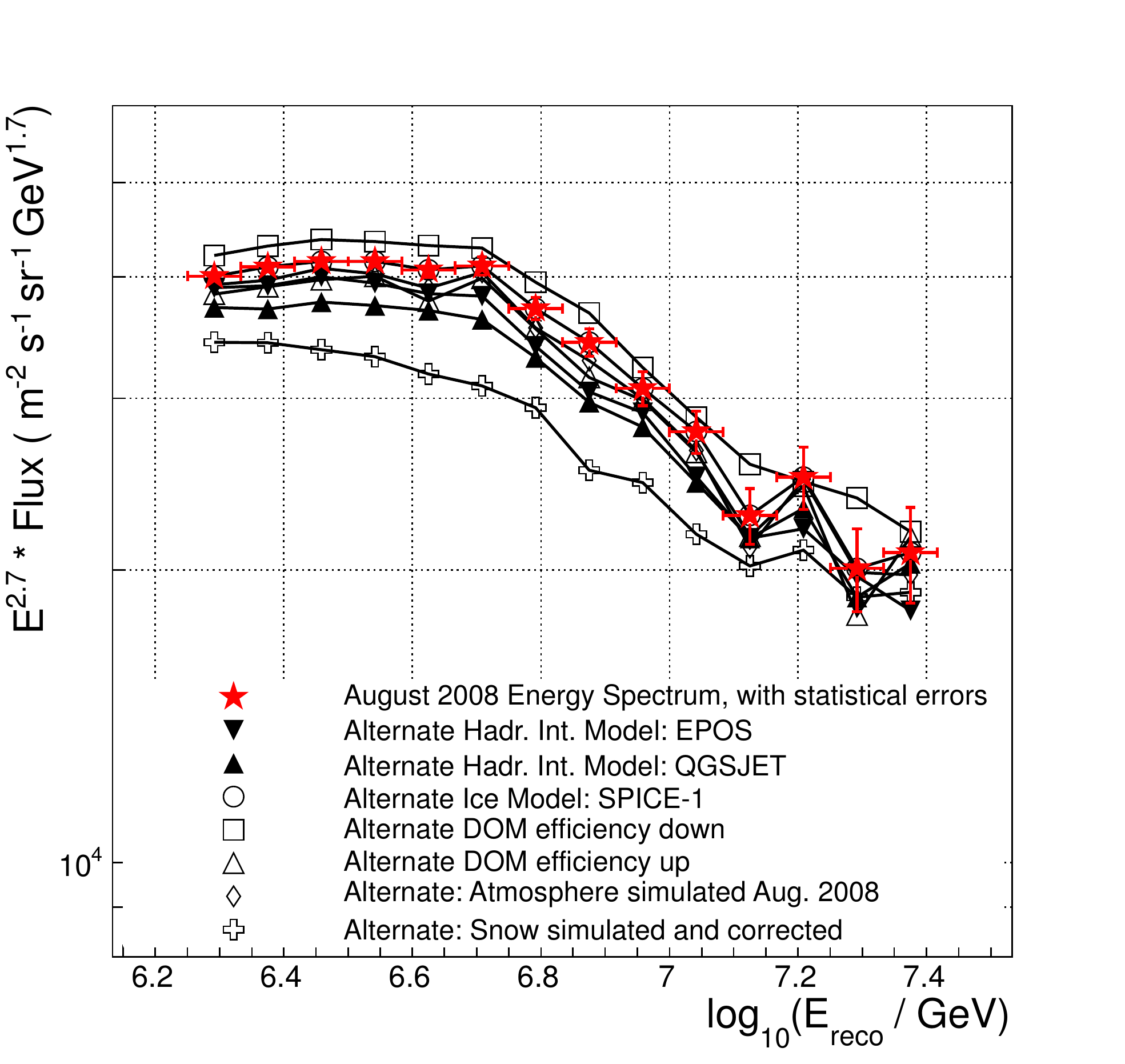}
\label{data_spectrum_output}
}
\caption[Aperture and Reconstructed Energy Spectrum from August 2008 Experimental Data]{\footnotesize 
{\it Upper left}: The efficiency (as defined in the text) with respect 
to the true energy, where (red) circles mark proton and (blue) squares denote iron.   
The efficiency is much smaller than 1: this is a combined effect of the requirement 
of coincidence between the two detectors, and the large radius within which the showers are generated.  
{\it Upper right}: The aperture with respect to the energy reconstructed by the neural network for all simulated data (all particle types), fitted as 
described in Eq.~(\ref{eqn:aperture_slanty}).
{\it Lower}: The reconstructed energy spectrum for the August 2008 data, which uses the aperture shown above.  
The additional symbols show alternate results from 
simulations using alternate models, as discussed in 
Section~\ref{sec:systematics}. }
\end{centering}
\end{figure}

The efficiency can be found in Fig.~\ref{eff_vs_e} for proton and iron separately.  To account for the fluctuations caused by the comparatively small statistics found in the simulated aperture distribution the aperture, $\mathcal{A}$, is fitted to a sigmoid function modified by a line, of the form:
\be
\mathcal{A} = \eta A \Omega = \frac{p_{0}}{1+e^{-p_{1}(E_0-p_{2})}} + \left(p_{3}E_0 + p_{4}\right),
\label{eqn:aperture_slanty}
\ee
as can be found in Fig.~\ref{ap_reco_slanty}, which shows the aperture for all generated particle types.  The slight loss of efficiency at the highest energies 
is due to events for which the likelihood minimization algorithm does not converge.
The resulting measured spectrum is shown in Fig.~\ref{data_spectrum_output}, where the minimum energy shown corresponds to the point of maximum efficiency from the fit Eq.~(\ref{eqn:aperture_slanty}), and the maximum shown is chosen to be two energy bins below the maximum energy simulated to avoid edge effects.

A phenomenological evaluation was developed \cite{Horandel:2003, Antonyan:2000}, 
where the differential cosmic ray flux was expressed as:
\begin{equation}
\frac{d\Phi_Z}{dE_0} (E_0)=\Phi^0_ZE_0^{\gamma_Z}\left[ 1+ \left(\frac{E_0}{\hat{E}_Z}\right)^{\epsilon_c}\right]^{\frac{\gamma_c -\gamma_Z}{\epsilon_c}},
\label{Eqn:hoerandel_flux}
\end{equation}
where $Z$ is the charge of a particle, $\Phi^0_Z$ is the absolute flux at 1 TeV, 
$\gamma_Z$ and $\gamma_c$ are the indices below and above the cut-off energy $\hat{E}_Z$, respectively, 
and $\epsilon_c$ is a parameter defining the curvature of the knee (where the knee becomes sharper as $\epsilon_c$ increases).  
The measured spectrum in Fig.~\ref{data_spectrum_output} has been fit to the 
 flux model in Eq.~(\ref{Eqn:hoerandel_flux}), with the resulting power law indices of 2.62 $\pm$ 0.05 at energies below and 3.26 $\pm$ 0.06 above a slowly turning knee around 4.98 $\pm$ 0.37~PeV, where the errors given are statistical.  The data points are given in Table~\ref{table:flux}, and the fit parameters are shown in Table~\ref{table:fit_params}.  The systematic error bars will be discussed in Section~\ref{sec:systematics}.

\begin{table}[htdp]
\begin{center}
\begin{tabular}{ccc}
\toprule
 Energy   &  $dN/dE$   $\pm$ stat $\pm$ sys  & Number of events \\ [1.5mm]
($10^6$~GeV) &  (GeV$^{-1}$ m$^{-2}$ s$^{-1}$ sr$^{-1}$) & \\ [1.5mm]
\hline\\ 
$	1.96	$	&	$(	4.12	\pm	0.04~^{+0.21}_{-0.60}	)$		$	\times 10^{-	13	}	$&	9642\\ [1.5mm]
$	2.37	$	&	$(	2.51	\pm	0.03~^{+0.13}_{-0.41}	)$		$	\times 10^{-	13	}	$&	6986\\ [1.5mm]
$	2.87	$	&	$(	1.52	\pm	0.02~^{+0.08}_{-0.29}	)$		$	\times 10^{-	13	}	$&	4948\\ [1.5mm]
$	3.48	$	&	$(	9.02	\pm	0.14~^{+0.43}_{-1.82}	)$		$	\times 10^{-	14	}	$&	3493\\ [1.5mm]
$	4.22	$	&	$(	5.27	\pm	0.10~^{+0.31}_{-1.15}	)$		$	\times 10^{-	14	}	$&	2401\\ [1.5mm]
$	5.11	$	&	$(	3.17	\pm	0.07~^{+0.14}_{-0.78}	)$		$	\times 10^{-	14	}	$&	1673\\ [1.5mm]
$	6.19	$	&	$(	1.71	\pm	0.04~^{+0.11}_{-0.36}	)$		$	\times 10^{-	14	}	$&	1138\\ [1.5mm]
$	7.50	$	&	$(	9.39	\pm	0.30~^{+0.67}_{-2.46}	)$		$	\times 10^{-	15	}	$&	703\\ [1.5mm]
$	9.09	$	&	$(	5.01	\pm	0.20~^{+0.25}_{-1.00}	)$		$	\times 10^{-	15	}	$&	489\\ [1.5mm]
$	11.01$	&	$(	2.69	\pm	0.14~^{+0.10}_{-0.58}	)$		$	\times 10^{-	15	}	$&	310\\ [1.5mm]
$	13.34$	&	$(	1.32	\pm	0.09~^{+0.17}_{-0.15}	)$		$	\times 10^{-	15	}	$&	206\\ [1.5mm]
$	16.16$	&	$(	8.59	\pm	0.64~^{+0.00}_{-1.37}	)$		$	\times 10^{-	16	}	$&	153\\ [1.5mm]
$	19.57$	&	$(	4.12	\pm	0.40~^{+0.74}_{-0.43}	)$		$	\times 10^{-	16	}	$&	98\\ [1.5mm]
$	23.71$	&	$(	2.55	\pm	0.29~^{+0.13}_{-0.33}	)$		$	\times 10^{-	16	}	$&	71\\ [1.5mm]
\bottomrule\\
\end{tabular}
\end{center}
\caption[]{\footnotesize The all-particle cosmic ray energy spectrum measured by the IceTop-IceCube 40-string/40-station arrays assuming hadronic interaction model SIBYLL-2.1.}
\label{table:flux}
\end{table}%

\begin{table}[htdp]
\begin{center}
\begin{tabular}{cc}
\toprule
Parameter &  Best Fit \\ [1.5mm]
\hline\\ 
$\Phi^0_Z$ / 10$^{-7}$m$^{-2}$s$^{-1}$sr$^{-1}$	&	$	1.81  \pm	0.65	$		\\ [1.5mm]
$E_0$ /  PeV 								&	$	4.75	\pm	0.59	$		\\ [1.5mm]  
$\gamma_z $								&	$	-2.61\pm	0.07 	$		\\ [1.5mm]  
$\gamma_c$								&	$	-3.23\pm	0.09	$		\\ [1.5mm]  
$\epsilon_c$								&	$	5.59	\pm	3.81	$		\\ [1.5mm]  
$\chi^{2}$/N$_{df}$							&	$	4.95	/ 9 $			\\ [1.5mm]  
\bottomrule\\
\end{tabular}
\end{center}
\caption[]{\footnotesize Fit parameters for the cosmic ray energy spectrum shown in Fig.~\ref{data_spectrum_output} using Eq.~(\ref{Eqn:hoerandel_flux}), as well as the chi-squared of the fit.  Statistical errors only are shown.}
\label{table:fit_params}
\end{table}%

\begin{table}[htdp]
\begin{center}
\begin{tabular}{cccccc}
\toprule
Energy Range	&	Experimental	&	\multicolumn{4}{c}{Monte Carlo Simulation}\\
\cmidrule{3-6}
(log$_{10}$(E$_{\mathrm{reco}}$ /GeV)) & Data & Protons & Helium & Oxygen & Iron		\\
\midrule
6.3-6.5 	&	18533	&	150	&	134	&	150	&	150	\\[1.5mm]
6.5-6.7 	&	8450 	&	165	&	142	&	132	&	133	\\[1.5mm]
6.7-6.9 	&	3461		&	125	&	174	&	134	&	151	\\[1.5mm]
6.9-7.1 	&	1214 	&	121	&	150	&	130	&	158	\\[1.5mm]
7.1-7.3 	&	427		&	148	&	144	&	147	&	152	\\[1.5mm]
7.3-7.5 	&	150		&	112	&	124	&	151	&	143	\\[1.5mm]
\bottomrule\\
\end{tabular}
\end{center}
\caption[]{\footnotesize Number of events of experimental data and simulation, for the six energy ranges analyzed for composition,  
after application of all cuts (including the selection for the newer half of the array).}
\label{table:nevents_finaleslices}
\end{table}%

\section{Composition}
\label{sec:comp}

As mentioned in Section~\ref{sec:nn}, mass composition requires a further minimization step.  Within each 
slice in energy and for each simulated species, the 
neural network produces a distribution of mass outputs which are referred to as
``template histograms''.  Examples of template histograms for all five species
are shown in Fig.~\ref{massoutput_alltypes}(right), for a particular slice in energy selected from
Fig.~\ref{massoutput_alltypes}(left).  Experimental data, when passed through the same neural network, also has a histogram of mass outputs which can be decomposed into a linear combination of the template histograms of the individual nuclear species (proton, oxygen, etc.).  A minimizer finds the optimal mixture of simulated species to match the experimental data.   

Note that the distributions of the templates in Fig.~\ref{massoutput_alltypes}(right) very rarely approach either 0 or 1 on the output axis: this is due to the overlap in the distributions, as shown in violet in Fig.~\ref{berries_eps}.  The type output from the network can be seen as a probability of being a certain species of particle: as the separation between species is never complete the probability that a particle is any given type will never be 1.  Furthermore, the higher mass particles have much more overlap between types, thereby causing probability of reconstructing an iron particle to be slightly lower.

To find the best mixture of species, the contribution of each species is quantified as a fraction, with the constraints
that each fraction must be between 0 and 1, and the fractions of all
nuclei must sum to unity.  The template histograms $C$ for each species are weighted
according to these fractions and summed to produce a histogram of their mixture 
$C_{mix}$.
Two species ``A'' and ``B''
(i.e. proton and iron only, as shown in Fig.~\ref{massoutput_2types}), are mixed using:
\begin{eqnarray}
\label{linear_comb}
C_{\mathrm{mix}} &=&  (f_A)C_A +(1 - f_A)C_B, 
\end{eqnarray}
where $f_A$ is the fraction of the mixture which is species $A$.  
Expanding to three species ``A'', ``B'', and ``C'', the mixture histogram is formed by:
\begin{eqnarray}
C_{\mathrm{mix}}  &=&  (f_A)C_A + (1 - f_A)\left[(f_B)C_B + (1 - f_B)C_C\right].
\label {eqn:phefe_fit}
\end{eqnarray}
Here, $f_A$ is the fraction which is species $A$, $f_B$ is the fraction of what remains
(which is not $A$) which is of species $B$. 
Mixtures of four or more species can be constructed with similar logic.
A minimizer can vary $f_A$, $f_B$, etc.~as free parameters.
Constructing the mixture in this way makes it easy to enforce in the minimizer that 
all fractions are physical (that is, between 0 and 1, and properly normalized).

The mixture of species that best matches the data is found using a $\chi^2$ test
and the software package \textsc{MINUIT} \cite{MINUIT:1975}, where:
\be
\chi^2 = \sum_{\mathrm{j=1}}^{\mathrm{all~bins}}\frac{(C_{\mathrm{data},j} - C_{\mathrm{mix},j})^2}{\sigma^2_{\mathrm{data},j}}.
\ee
Here, $C_{\mathrm{data}}$ is the measured data histogram, as reconstructed by the neural network output, and $\sigma^2_{\mathrm{data}}$ is the variance in the data.  $C_{\mathrm{mix}}$ is a mixture of simulated species as described above.  
An example of a best-fit mixture of template histograms for a slice in energy using only two species is shown in Fig.~\ref{massoutput_2types}, where the 
"testing'' sample from Section~\ref{sec:nn} (including all species) was treated as though it were data, and compared to 
template histograms of proton and iron only (for easier viewing).  

\begin{figure}[tbp]
\begin{centering}
\subfigure[]{
\includegraphics[width=.9\textwidth]{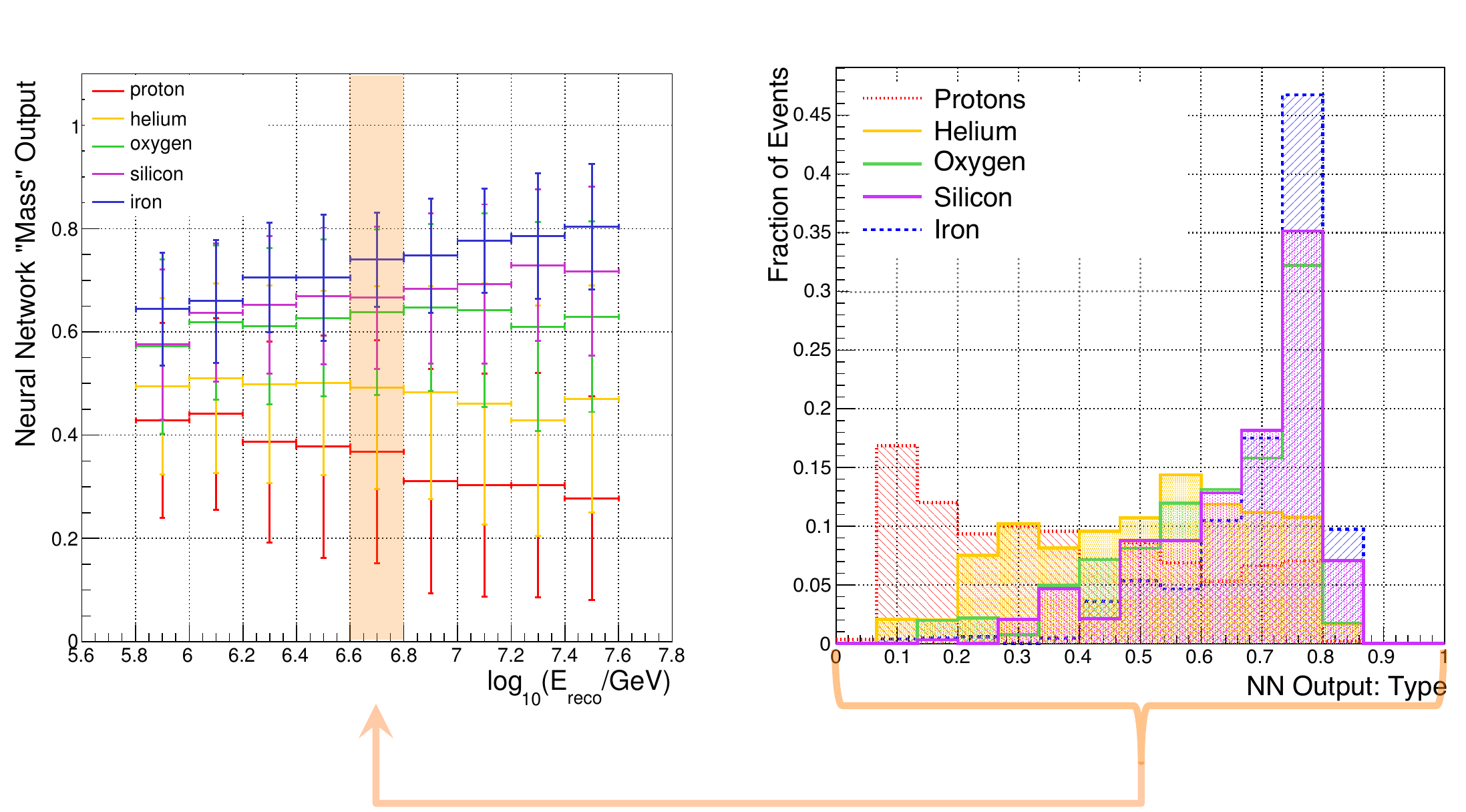}
\label{massoutput_alltypes}
}
\subfigure[]{
\includegraphics[width=.5\textwidth]{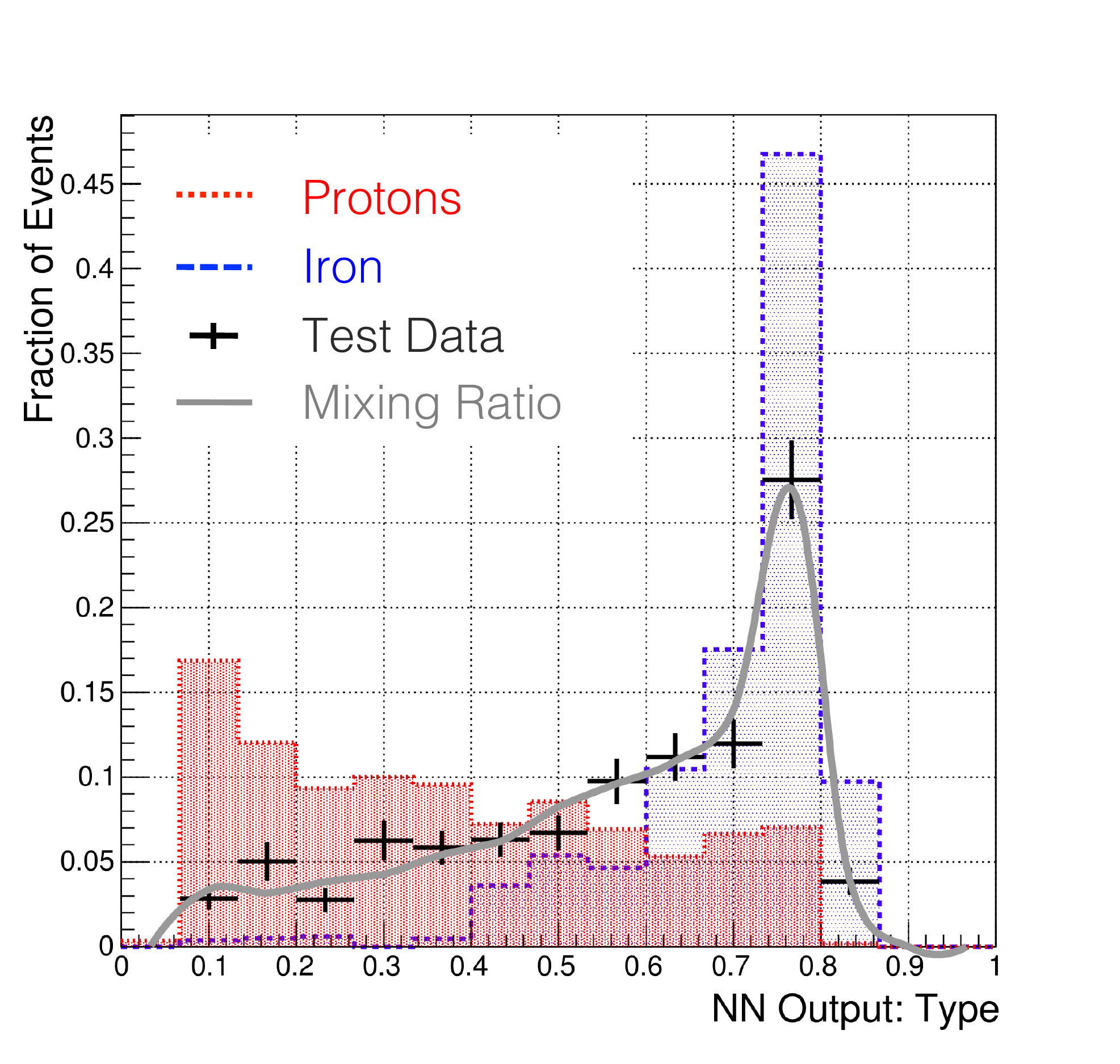}
\label{massoutput_2types}
}
\caption[template histograms for All Particle Types for Minimization Across One Slice in Energy]{\footnotesize 
(a), {\it left}: the mass output from the neural network is shown as RMS profiles in slices of the reconstructed energy with colors corresponding to the different species, as labeled.  This plot is analogous to Fig.~\ref{berries_eps} after the neural network transformation to the new axes; therefore, the large overlap at the lowest energies corresponds to the similarly large overlap in Fig.~\ref{berries_eps}.  
(a), {\it right}: The underlying distributions, or template histograms, for a single slice in energy 
for all 5 particle species.  
(b): An example of template histogram fitting: the testing simulation, or ``test data'', sample (black with statistical error bars), which includes all five species, was fit to the proton and iron template histograms only -- for easier viewing -- using the minimization technique described in the text.  The grey ``mixing ratio'', or combination of proton and iron comprising the test data, matches the data quite well even though only two template histograms were used to fit a sample containing five species.}
\label{mini_step0_5type}
\end{centering}
\end{figure}

Simulated nuclei from the testing sample (described in Section~\ref{sec:nn}) were ``hand-mixed'' in different proportions at different energies to test 
the effectiveness of various minimization strategies, ranging from using only the two most extreme template histograms (protons and iron) 
with only one free parameter, to using all five template histograms with four free parameters.  With too few nuclei, the minimizer is confident about
the best mixture, but the mixture is not a very good fit.  With too many, the minimizer can find many options for achieving a good fit (many of
them unphysical mixtures, such as iron and oxygen present with no silicon in between), so it is less confident about the best one.
Using three template histograms: protons, ``intermediate nuclei'' (a 50-50 mixture of helium and oxygen), and iron,
is a balanced approach:  
\begin{equation}
\label{linear_comb_ohe}
C_{mix} = f_{p}C_p +(1-f_{p})\left[(f_{fe})C_{fe} + (1-f_{fe})(0.5 \cdot C_o + 0.5 \cdot C_{he})\right]. \\
\end{equation}
A list of the number of events for each species as a function of energy is given in Table~\ref{table:nevents_finaleslices}.  The mass output distributions with respect to energy are shown for the final simulated ``analysis'' sample in Fig.~\ref{mini_result_pheofe_separate}.  The results of a check of this method using a mixed composition sub-set of the ``testing'' sample of simulated data are shown in Fig.~\ref{mini_result_pheofe_together}, where the true mass value (gold bars) is compared with the reconstructed mass value (red stars).  
This minimization technique reconstructs the true mean logarithmic mass within the error bars for each slice in energy.  Furthermore, Fig.~\ref{mini_result_pheofe_protons} and Fig.~\ref{mini_result_pheofe_iron} show this technique applied to pure proton and iron subsets of the ``testing'' sample of simulated data, respectively, to demonstrate the capability of the method to reconstruct both extremes.   As discussed above, as the probability of finding an iron is never 1, the iron reconstruction never quite reaches the expected value but remains within statistical error bars.

\begin{table}[tdp]
\begin{center}
\begin{tabular}{cc}
\toprule
Energy Range	&	Reconstructed \\
(log$_{10}$(E$_{\mathrm{reco}}$ /GeV)) & \lna $\pm$ stat. $\pm$ sys.\\
\midrule
6.3-6.5 	& 	$2.17 \pm  0.03 ~^{+0.38}_{-0.38}$\\ [1.5mm]
6.5-6.7 	& 	$2.28 \pm  0.06 ~^{+0.55}_{-0.37}$\\ [1.5mm]
6.7-6.9 	& 	$2.50 \pm  0.09 ~^{+0.59}_{-0.33}$\\ [1.5mm]
6.9-7.1 	& 	$2.86 \pm  0.14 ~^{+0.65}_{-0.34}$\\ [1.5mm]
7.1-7.3 	& 	$3.36 \pm  0.16 ~^{+0.31}_{-0.32}$\\ [1.5mm]
7.3-7.5 	& 	$3.48 \pm  0.23 ~^{+0.30}_{-0.15}$\\ [1.5mm]
\bottomrule\\
\end{tabular}
\caption[]{\footnotesize \lna~for six analyzed energy ranges.
The subscripts (superscripts) are the minimum (maximum) alternate \lna~values, 
found by passing data (with \K{} and \s{} adjusted according to alternate models)
through the neural network and the minimization procedure.}
\label{table:lna}
\end{center}
\end{table}%

\begin{figure}[htbp]
\begin{centering}
\subfigure[]{
\includegraphics[width=.45\textwidth]{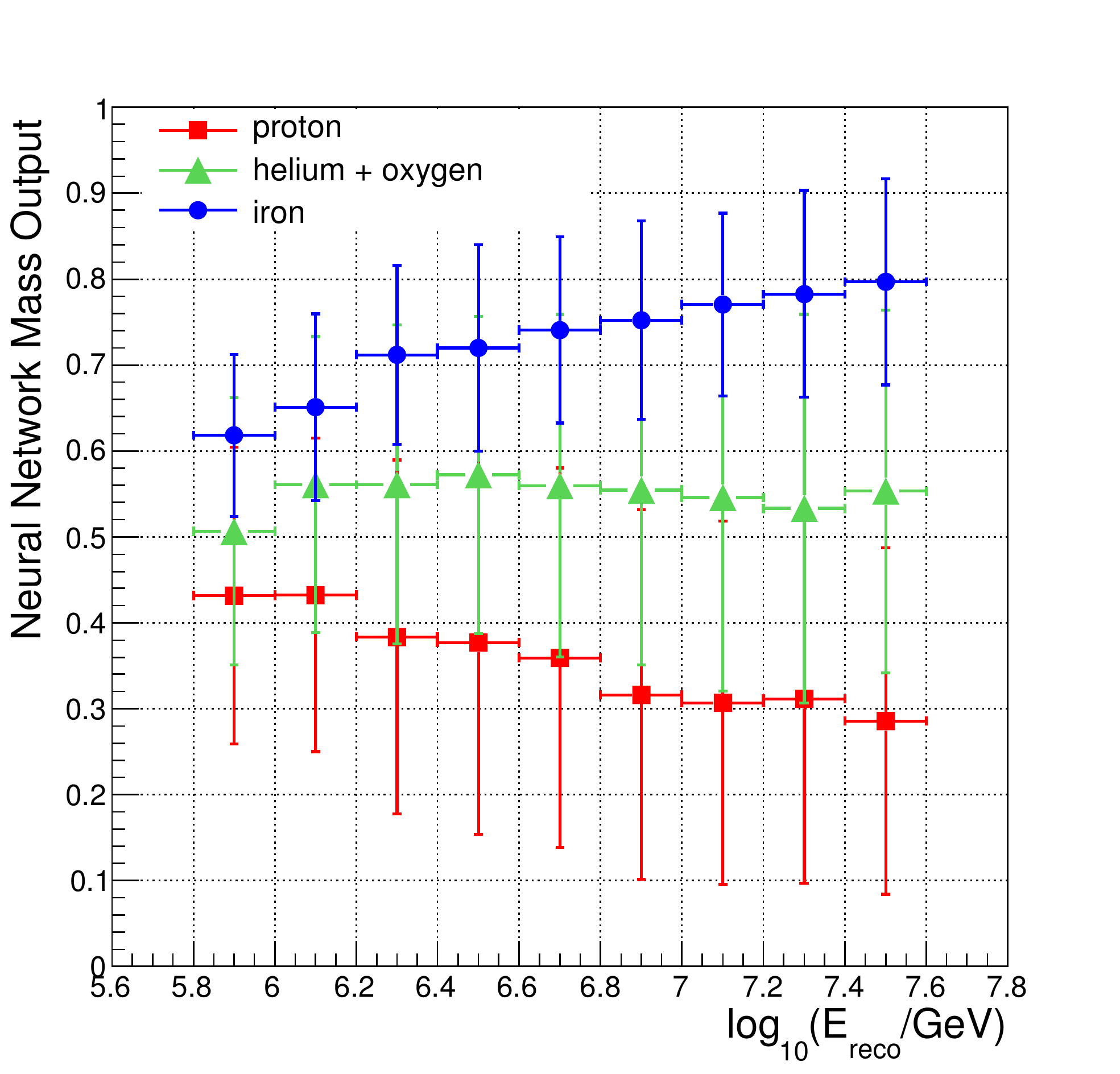}
\label{mini_result_pheofe_separate}
}
\subfigure[]{
\includegraphics[width=0.45\textwidth]{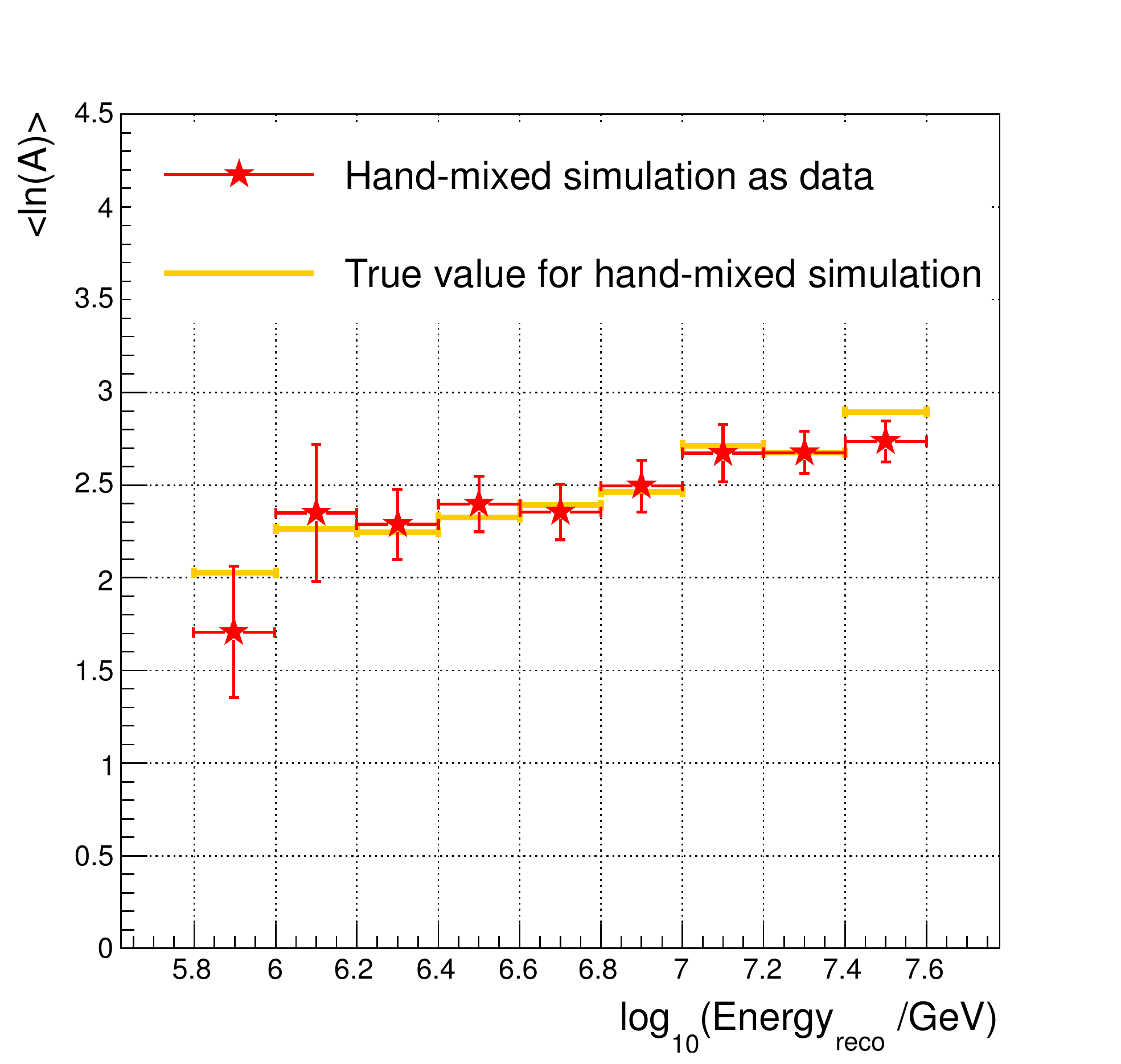}
\label{mini_result_pheofe_together}
}
\subfigure[]{
\includegraphics[width=0.45\textwidth]{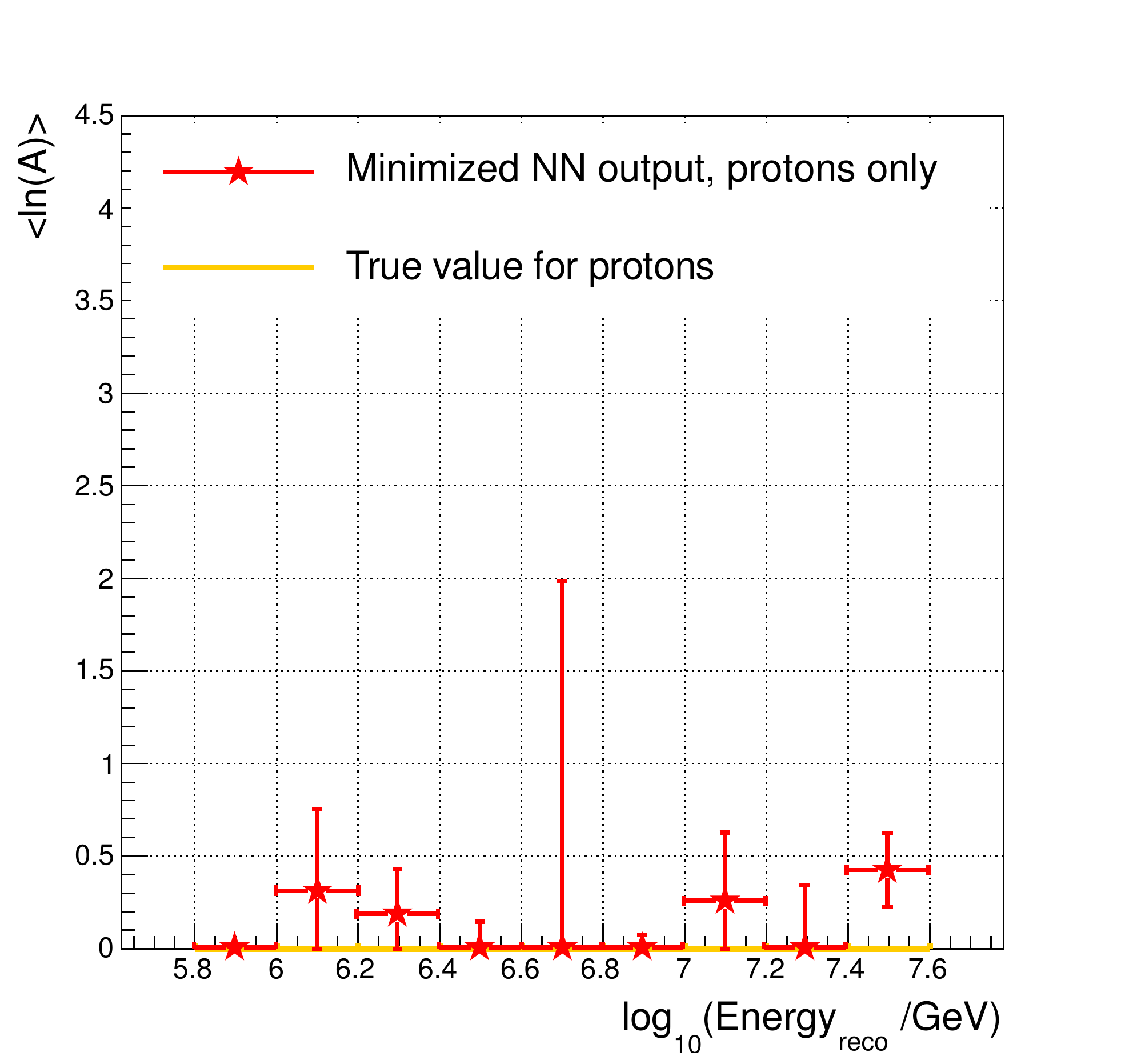}
\label{mini_result_pheofe_protons}
}
\subfigure[]{
\includegraphics[width=0.45\textwidth]{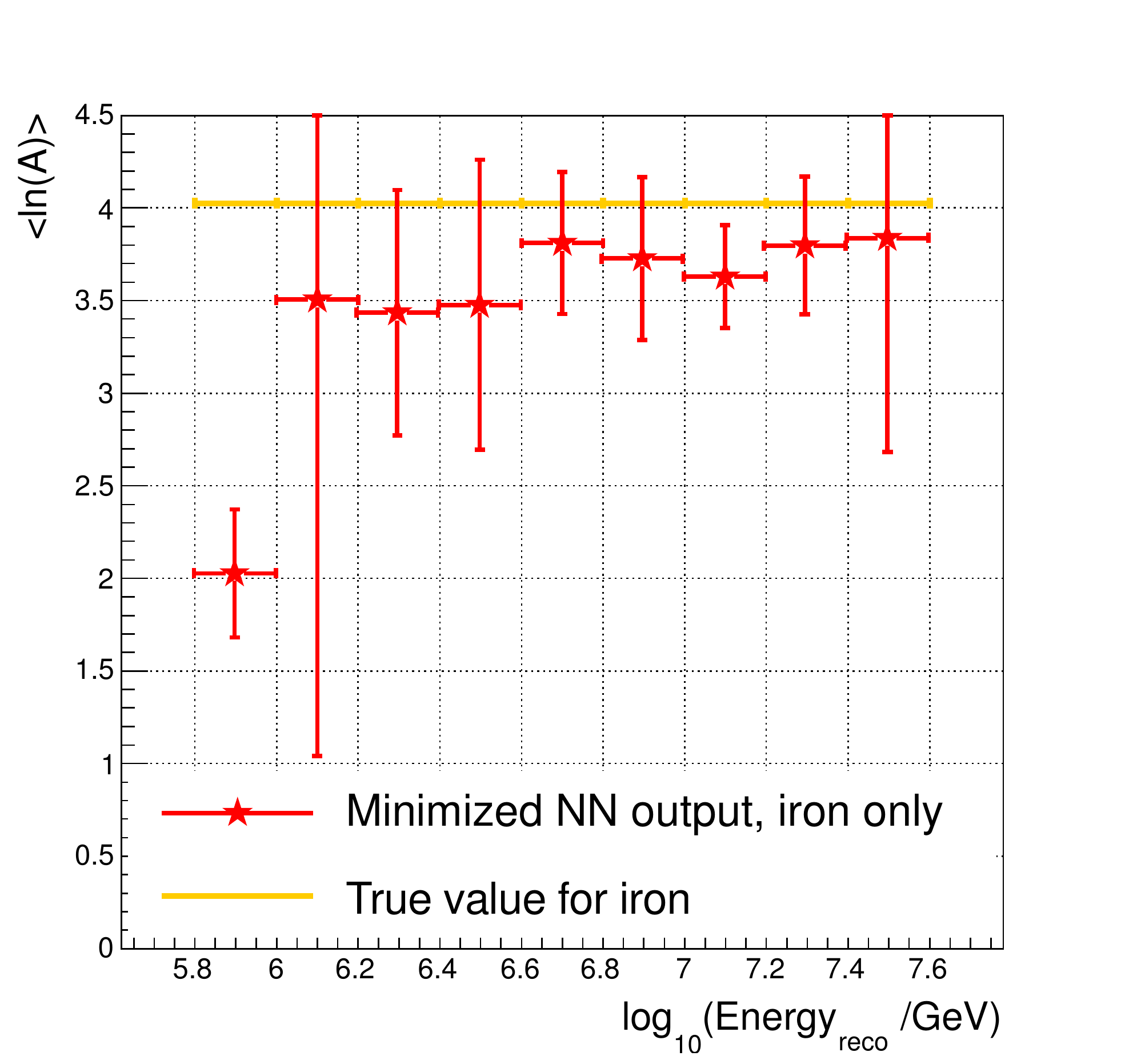}
\label{mini_result_pheofe_iron}
}
\caption[Result of 3-type minimization on testing sample for all energies]{\footnotesize Performance of mass reconstruction.  (a): profiles of mass output vs reconstructed energy for the chosen simulated species used in the minimization: proton, helium/oxygen mixture, and iron.  The parameter space is well-covered by these distributions, as demonstrated by the overlap.  (b): The result of a check of the 3-type (p/He-O/Fe) minimization scheme for all energies using a sub-set of the ``testing'' sample of simulated data.  The true \lna~for each slice in energy is marked in gold, while the \lna~chosen by the minimizer as the best match for the hand-mixed simulation is shown in red, with statistical error bars.  Even with this small testing sample the minimizer finds a very good fit to the input values at energies where the detector is fully efficient (i.e. above 6.2 in log$_{10}$(E$_\mathrm{reco}$)).  Similarly (c) and (d) show the ability of this technique to reconstruct pure proton and pure iron samples respectively at the proper mass.}
\label{mini_result_pheofe}
\end{centering}
\end{figure}

The composition results of this combined neural network and minimization technique applied to the August 2008 data are shown in Fig.~\ref{comp_final_syserrors}, with data points and statistical error bars in red.  The systematic errors (shown as alternate
symbols) are discussed below and are summarized in Table~\ref{table:lna}.

\begin{figure}[htbp]
\begin{centering}
\includegraphics[width=.5\textwidth]{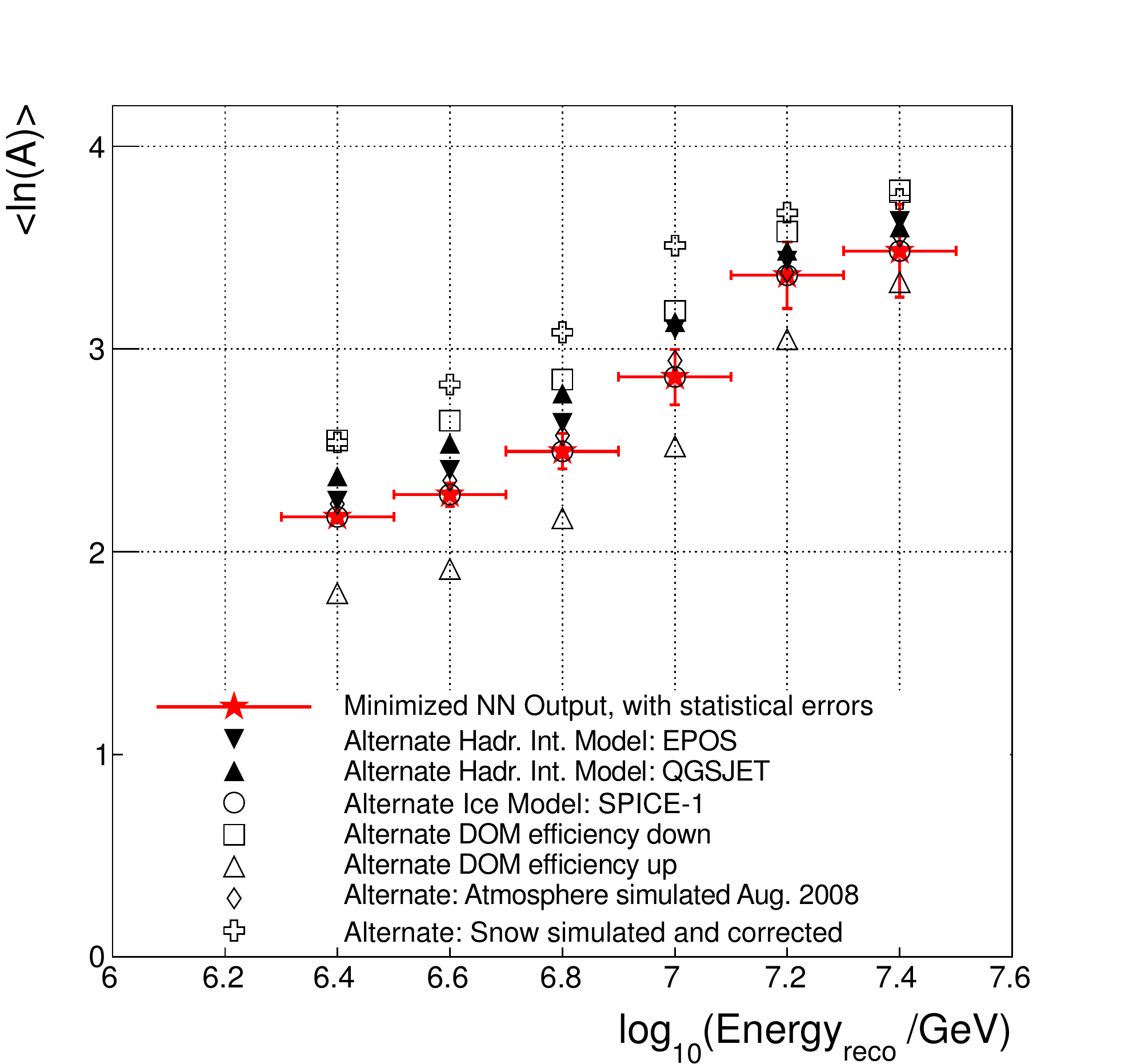}
  \caption[Composition Results With Systematic Errors]{\footnotesize The mean logarithmic mass vs primary energy for August 2008 data as calculated using the analysis technique described in this work.  The error bars on the data points are statistical, 
while the additional symbols indicate results after systematic shifts estimated using alternate models.}
\label{comp_final_syserrors}
\end{centering}
\end{figure}

\section{Systematic Uncertainties}
\label{sec:systematics}
This measurement of cosmic ray composition and energy spectrum depends crucially on the two independent observables \K{} and \s{}.
Systematic errors, particularly in the scaling of one observable 
(independent of the other) can make the composition seem heavier or lighter.
To estimate the magnitude of potential systematic errors and their effect on
the measurement of composition,
the \K{}-\s{}~parameter space (as depicted in Fig.~\ref{berries_eps}) was studied using samples of simulated events with alternate
models, listed below.  

\begin{itemize*}
\item{\it Hadronic Interaction Model:}
As the forward-direction physics of the first interaction of high-energy cosmic rays
has not yet been measured by accelerator experiments on Earth, 
cosmic ray physicists depend upon extrapolations from the known
physics at lower energies.
There are, therefore, a number of different high-energy hadronic interaction models. 
\emph{Model used:} SIBYLL-2.1 \cite{Ahn:2009}. 
\emph{Alternate models:} EPOS-1.99 \cite{EPOS:2007} and QGSJET-II-03 \cite{qgsjet:1997}.  
\item {\it Optical Properties of the Ice:} 
For the 40-string configuration of IceCube there were two models available for measuring the depth and 
wavelength-dependence of the scattering and absorption in the ice.  In general, the absorption lengths 
vary from 100~m to 200~m and effective scattering lengths range from 20~m to 70~m due to the concentration 
of dust in the glacial ice, which has a strong depth-dependence \cite{IceModel:2006,Chirkin:2011,Price:2000}.  
\emph{Model used:} AHA \cite{IceModel:2006}, the standard ice model for the 40-string configuration of IceCube and was 
developed using data from the predecessor of IceCube (AMANDA).  
\emph{Alternate model:} SPICE-1, which measured the ice properties over the full depth range of 
IceCube using the in-situ LEDs present on every DOM main~board \cite{Chirkin:2011}. 
\item{\it DOM Efficiency:} 
The efficiency of the DOMs in the ice depends upon a number of factors including the PMT quantum efficiency, the transmission of the glass and the gel, and the refrozen ice from the deployment hole (which is less clear than the bulk ice).  Measurements of the DOM efficiency in a controlled setting have revealed an estimated uncertainty of approximately $8\%$ \cite{Abbasi:2010a}, which is equivalent to a shift in log$_{10}$(\K{})  of $0.035$ in log$_{10}$(\K{}).
This was treated as a constant systematic 
uncertainty across all values of \s{} which could either raise or lower the measured \K{} in IceCube.
IceTop is not affected because its signals are calibrated to Vertical Equivalent Muons, as described in 
Section~\ref{sec:intro}.
\item{\it Snow correction in reconstruction:}
As discussed in Section~\ref{sec:snow}, snow accumulation at the surface can affect signals in IceTop, and reconstruction 
methods can correct for this effect in data.  However, the correction applied during reconstruction will not work perfectly
for all events at all energies.  Potential systematic errors in \s{} due to imprecision in the technique itself were studied 
by comparing simulations in which snow was "removed" during reconstruction to simulations in which there was no snow at all.
\item{\it Atmosphere: effect on high-energy muons:}
Atmospheric conditions high in the atmosphere, and the effective temperature of the atmosphere,
affect the number of high-energy muons and \K{}. 
With the selection of one month of data for this analysis, the effective temperature was stable; K70
exhibits negligible daily or weekly variation within that month.  In addition,
the atmospheric profile high in the atmosphere was compared
to that simulated by CORSIKA, and the systematic error due to this
mismatch is also negligible compared to other systematic effects on \K{} discussed in this
section.   
\item{\it Atmosphere: effect on electromagnetic surface component:}
Atmospheric conditions near the
surface (such as pressure or density) affect the electromagnetic component measured by \s{}.
The actual average atmospheric conditions during August 2008 differs from that 
simulated in CORSIKA.  The systematic effect on \s{} due to this mismatch was estimated by
comparing CORSIKA showers (10~PeV and 10~degree zenith angle) made using the standard South Pole winter atmosphere profile
to those made using a customized profile built from measurements taken during August 2008. 
Distributions of $\log_{10}(S125)$ differ by less than 0.01, and this potential shift was treated as
constant across all energies. 
Understanding these effects in IceTop and IceCube is the subject of ongoing study.

\end{itemize*}

Some of these effects (such as ice model and DOM efficiency) affect only the in-ice measurement of \K{},
while others (such as snow removal and atmospheric pressure) affect only the surface measurement of \s{}.
The hadronic interaction model potentially affects both.
A summary of how \K{} and \s{} are observed to shift when alternate models are explored
is shown in Fig.~\ref{systematics_curves} on the left and right, respectively.
Potential shifts in log$_{10}$(\K) are shown as a function of log$_{10}$(\s), while potential
shifts in log$_{10}$(\s) are shown as a function of log$_{10}$(\K).

To explore how the final results (composition and spectrum) would change under an alternate model,
values of either log$_{10}$(\K) or log$_{10}$(\s) in the data were artificially shifted according to the estimated curves in Fig.~\ref{systematics_curves}.
(For alternate hadronic interaction models, events were shifted half in \K{} and half in \s{}.)
For each alternate model, these shifted data sets were then 
run through the full analysis machinery (neural network reconstruction of $A$ and $E_0$ and minimization with template
histograms) to yield "alternate" composition and energy spectrum results.  
Although primitive, this procedure allowed for exploration of many alternate models quickly, without having to generate
large sets of simulated data for each one.

Figure~\ref{comp_final_syserrors} shows the composition result (with standard simulations) together with these alternate
results, after applying a shift either to \K{} or to \s{} (or both).  
These results, including errors, are also summarized numerically in Table~\ref{table:lna}.  
The "systematic errors" are drawn from the maximum and minimum \lna yielded by any of the alternate models
at each energy.
Figure~\ref{data_spectrum_output} shows a similar collection of alternate results for the energy spectrum.
The numeric values are given in Table~\ref{table:flux}.

\begin{figure}[htbp]
\begin{centering}
\includegraphics[width=\textwidth]{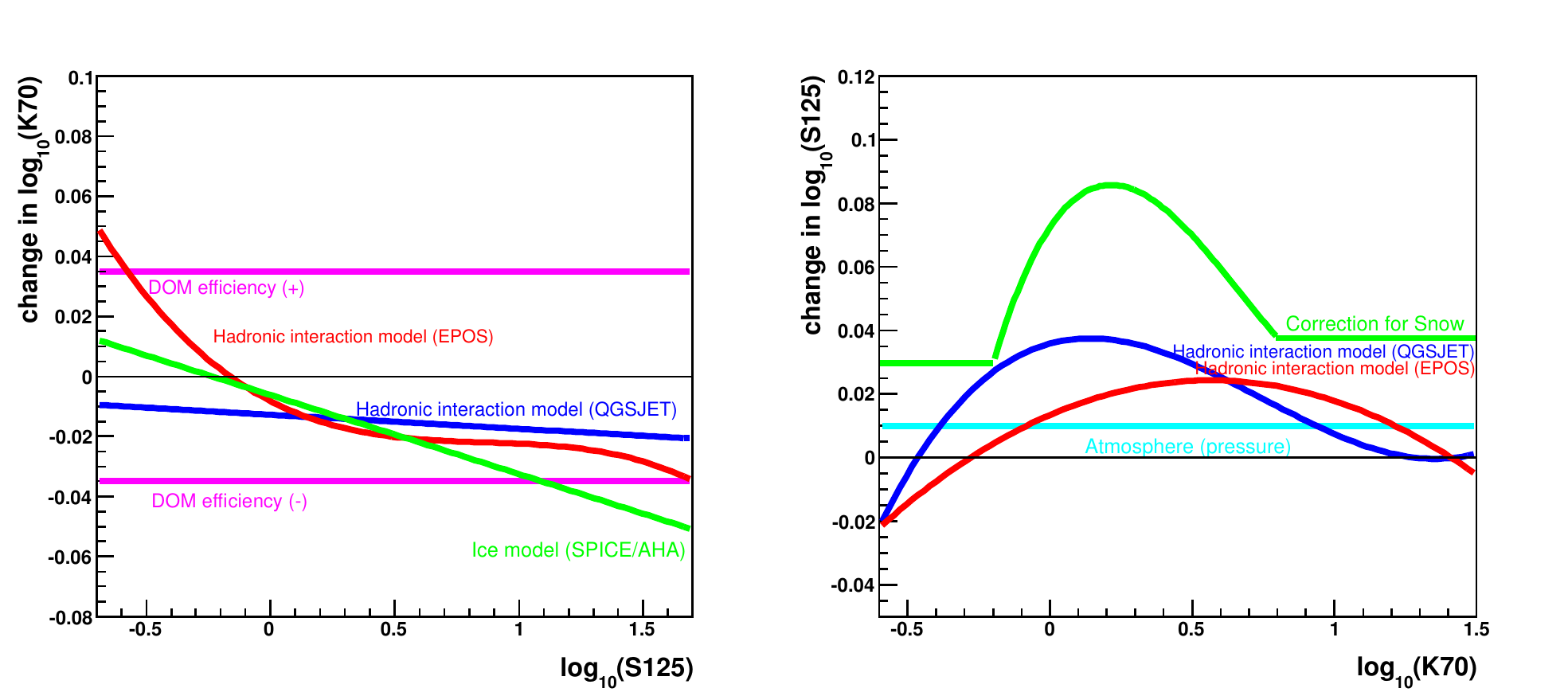}
\caption[Systematic errors]{\footnotesize {\it Left}: Estimates of systematic uncertainties affecting the in-ice measurement
(as shifts in log$_{10}$\K{} for a given log$_{10}$\s).  
{\it Right}: Estimates of systematic uncertainties affecting the surface measurement
(as shifts in log$_{10}$\s{} for a given log$_{10}$\K).}
\label{systematics_curves}
\end{centering}
\end{figure}

\section{Results and Discussion}

\begin{figure}[htbp]
\begin{centering}
\subfigure[]{
\includegraphics[width=.75\textwidth]{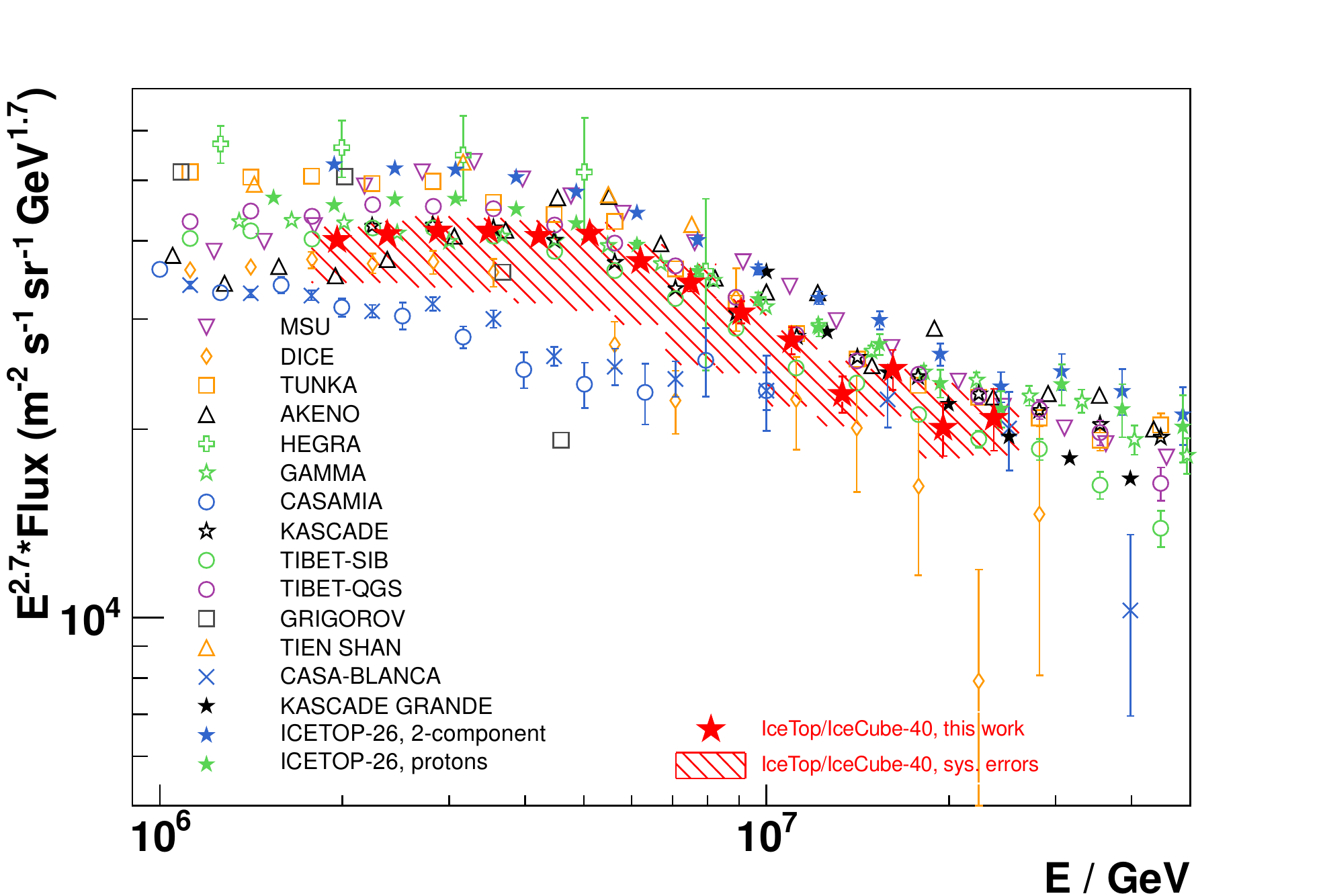}
\label{espec_almostall}
}
\subfigure[]{
\includegraphics[width=.75\textwidth]{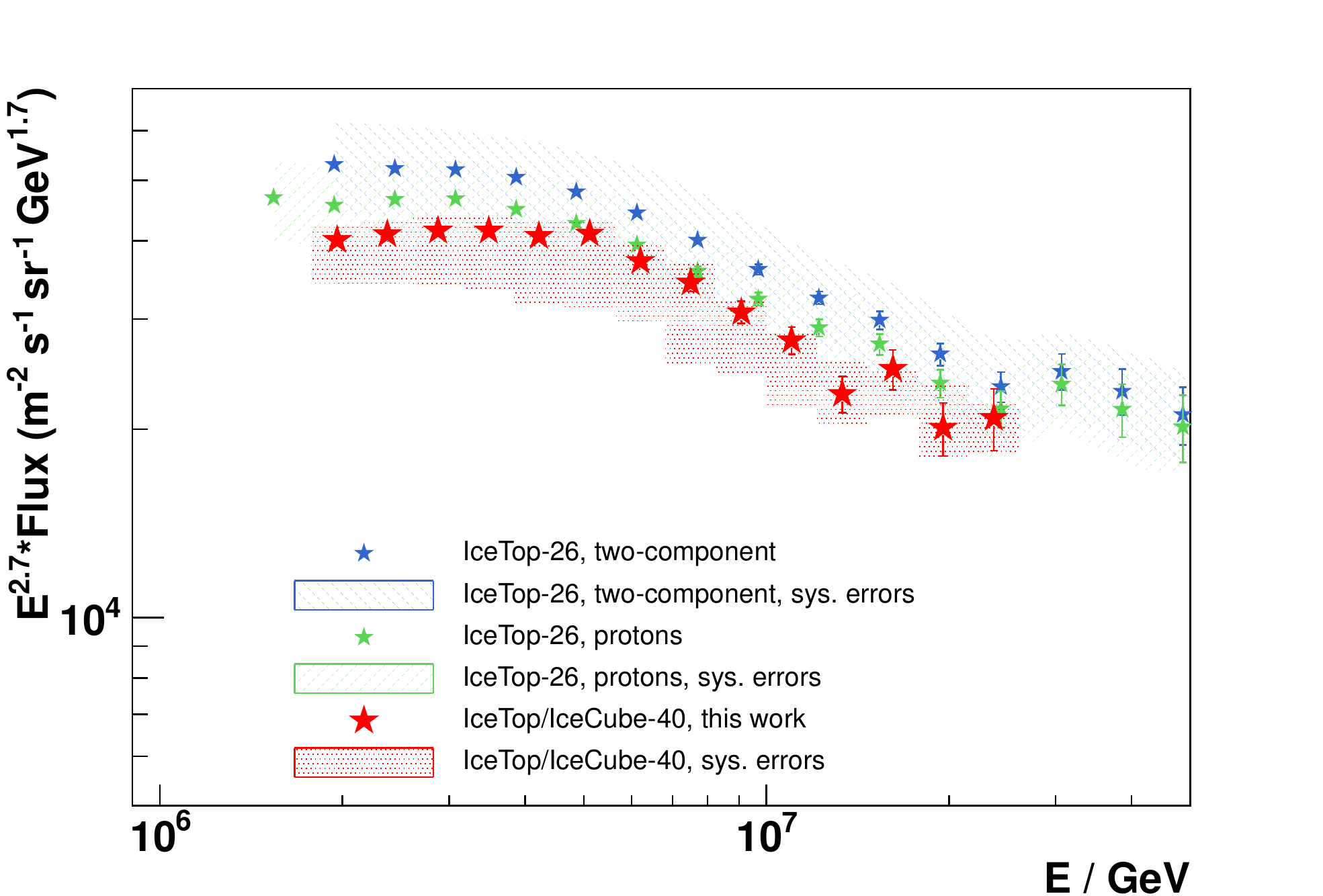}
\label{espec_it26only}
}
\caption[All Particle Energy Spectrum]{\footnotesize The all-particle
cosmic ray spectrum found using data from 40-station IceTop/IceCube as described in the text.  
For this analysis the results are shown as (red) stars, with statistical error bars as solid (red) lines and systematic errors as the shaded (red) region.  {\it Upper}: this result compared to other experiments with their statistical errors only.
{\it Lower}: this result compared to two spectra from IceTop-26, with systematic errors also shown.
Fluxes are multiplied by $E^{2.7}$, and the scales of the two plots are the same.  
The overall flux and shape found in this  work are very similar to those reported by other experiments.  Data points from \cite{Kislat:2011}.}
\label{espec_all}
\end{centering}
\end{figure}

\begin{figure}[htbp]
\begin{centering}
\includegraphics[width=.75\textwidth, angle=0]{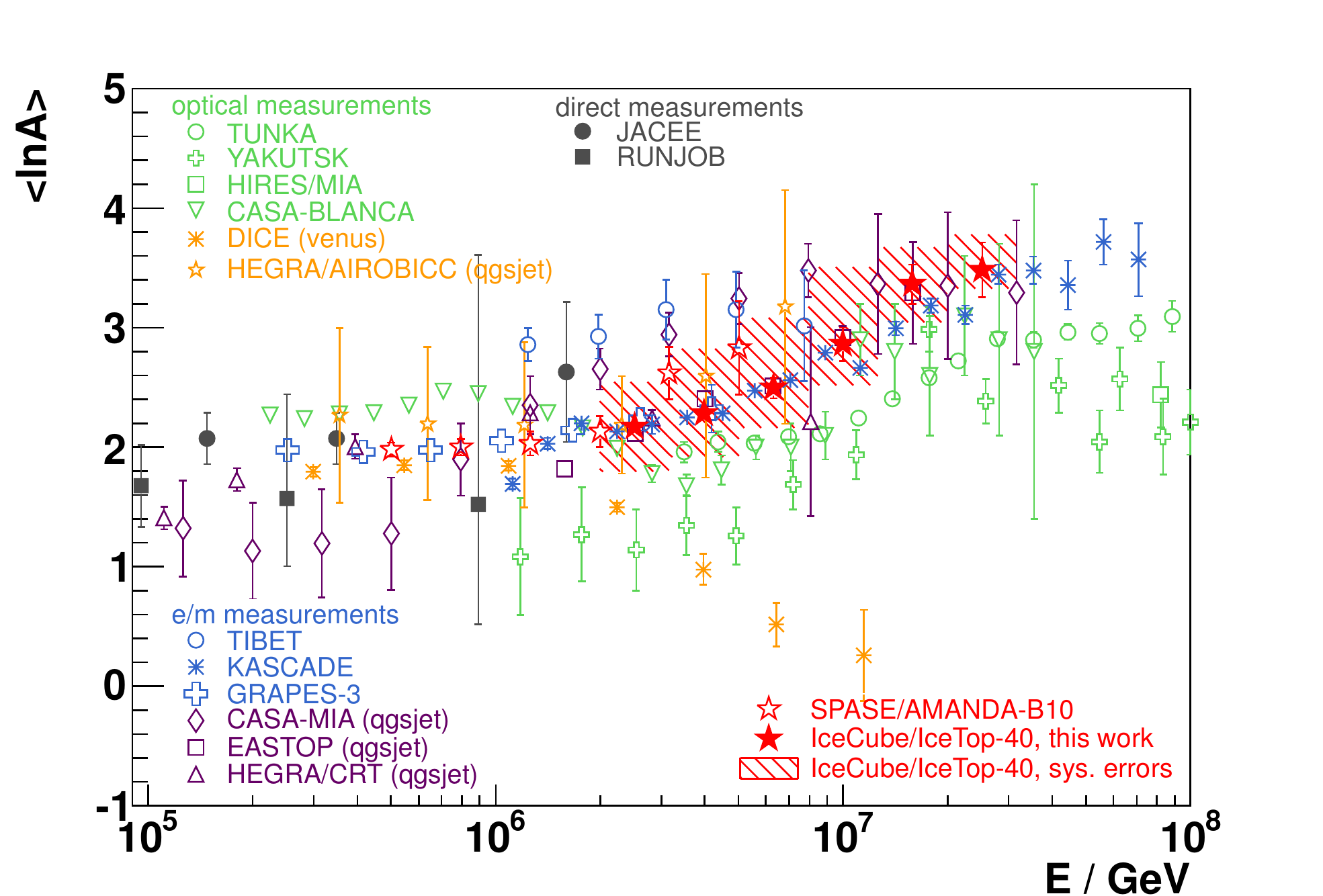}
\caption[Mean Logarithmic Mass]{\footnotesize The mean logarithmic mass vs
primary energy for a number of experiments, as labeled.  The optical and e/m measurements used the SIBYLL hadronic interaction model unless noted in parentheses.  The IceTop/IceCube 40-string/40-station results are shown in (red) stars, with solid (red) error bars indicating the statistical errors, while the shaded red region represents the systematic errors.  
The data indicate an increasing mass through the knee.
The results of this analysis are similar to measured results from some other experiments (in particular, most e/m experiments) as well as the 
flux model from Eq.~(\ref{Eqn:hoerandel_flux}), but dissimilar to others (in particular, optical measurements).  
Data points compiled from \cite{Horandel:2003,Kampert:2012}. }
\label{comp_all}
\end{centering}
\end{figure}

Data from the combined IceTop and IceCube detectors in their 2008
40-string/40-station configuration have been used to develop a 
technique to measure the cosmic ray energy spectrum and composition at
energies between 1~PeV and 30~PeV.  Using measurements of the
electromagnetic component of the air shower at the surface (\s) and the
muonic component of the air shower in the ice (\K),
a neural network in conjunction with a $\chi^2$ minimization algorithm
have been employed to extract the relevant physical parameters:
\lna~and $E_0$.

This technique provided an energy resolution better than $14\%$ with small offset due to primary mass, which led to an energy spectrum that compares
favorably with other recent measurements shown in Fig.~\ref{espec_all}.   
In these figures, the shaded (red) region indicate the systematic errors for this analysis.  Comparisons are made to
other experiments (with statistical error bars only) in Fig.~\ref{espec_almostall}.
Figure~\ref{espec_it26only} highlights a comparison (including systematic errors) to two spectra measured by the 26-station configuration of IceTop only (IT-26) \cite{Kislat:2011}, 
one assuming protons and one assuming a two-component model.  
This work (using a coincidence detector and a neural network based on both surface and deep observables) and the IT-26 work
(using the IceTop surface detector only and an unfolding technique)
yield slightly different spectra, but are consistent within systematic errors.
When fit to the 
flux model in Eq.~(\ref{Eqn:hoerandel_flux}), the spectrum
presented here indicates a power law of indices 2.61 $\pm$ 0.07 below and 3.23 $\pm$ 0.09 
above a slowly turning knee around 4.75 $\pm$ 0.59~PeV, where the errors given are statistical.

This method also provided a reconstructed mass parameter and, in turn,
a measurement of cosmic ray composition: the mean logarithmic mass, \lna.  The
mass composition 
is compared with other experimental measurements in Fig.~\ref{comp_all} where, additionally, systematic error bars are shown for IceCube/IceTop.
In this analysis, using the 40-string/40-station configuration of IceTop and IceCube from 2008, a mass 
is observed which similar to some other measured results (especially those using 
a similar electons vs. muons stretegy as in this work), and in disagreement with others (especially the optical measurements, which emply a very different strategy).
The slope of the strong increase in mass through the knee region is nearly identical to the 
model in Eq.~(\ref{Eqn:hoerandel_flux}). 
Additionally, this new analysis
technique with IceCube-40/IceTop-40 shows results consistent within
systematic errors to those from a different technique applied to data from its
predecessor, SPASE-2/AMANDA-B10 \cite{Ahrens:2004}.  

In the future this technique can be expanded to include new composition-sensitive input parameters,
as well as employ the larger IceCube detector and a greater quantity of data.  
From looking at several simulated alternate models, it is clear that systematic uncertainties (both in the
in-ice measurement and surface measurement) can greatly affect the measured composition and spectrum,
regardless of the detector's size or livetime.
Although the mean logarithmic mass itself is difficult to measure absolutely because of systematics,
this work shows an unmistakable \emph{trend} of increasing mass, regardless of the differing absolute
scale of \lna~measured with respect to different models.
The systematic studies performed for this analysis have led to many improvements which will allow for more precise measurements with new data.  Therefore, in the coming years the complete IceCube Neutrino Observatory will be the only detector of its kind able to provide both composition and energy spectrum measurements from energies overlapping with direct measurements below the knee to energies nearing the ankle.

\section{Acknowledgements}
We acknowledge the support from the following agencies: U.S. National Science Foundation-Office of Polar Programs, U.S. National Science Foundation-Physics Division, University of Wisconsin Alumni Research Foundation, the Grid Laboratory Of Wisconsin (GLOW) grid infrastructure at the University of Wisconsin - Madison, the Open Science Grid (OSG) grid infrastructure; U.S. Department of Energy, and National Energy Research Scientific Computing Center, the Louisiana Optical Network Initiative (LONI) grid computing resources; National Science and Engineering Research Council of Canada; Swedish Research Council, Swedish Polar Research Secretariat, Swedish National Infrastructure for Computing (SNIC), and Knut and Alice Wallenberg Foundation, Sweden; German Ministry for Education and Research (BMBF), Deutsche Forschungsgemeinschaft (DFG), Research Department of Plasmas with Complex Interactions (Bochum), Germany; Fund for Scientific Research (FNRS-FWO), FWO Odysseus programme, Flanders Institute to encourage scientific and technological research in industry (IWT), Belgian Federal Science Policy Office (Belspo); University of Oxford, United Kingdom; Marsden Fund, New Zealand; Australian Research Council; Japan Society for Promotion of Science (JSPS); the Swiss National Science Foundation (SNSF), Switzerland. 

\bibliographystyle{phaip}
\bibliography{largebib}

\begin{thebibliography}{10}

\bibitem{Bell:1978}
A.~R. Bell,
\newblock Monthly Notices of the Royal Astronomical Society {\bf 182}, 443
  (1978).

\bibitem{Horandel:2009}
J.~Blumer, R.~Engel, and J.~R. H\"{o}randel,
\newblock Progress in Particle and Nuclear Physics {\bf 63}, 293 (2009).

\bibitem{Torres:2004}
D.~F. Torres and L.~A. Anchordoqui,
\newblock Reports on Progress in Physics {\bf 67}, 1663 (2004).

\bibitem{Stanev:2007}
T.~Stanev,
\newblock Proceedings of the 30th International Cosmic Ray Conference, Merida,
  Mexico {\bf arXiv:0711.2282v1} (2007).

\bibitem{Horandel:2004}
J.~H\"{o}randel,
\newblock Astroparticle Physics {\bf 21}, 241 (2004).

\bibitem{IceTopDetectorPaper:2012}
R.~Abbasi et~al.,
\newblock submitted to Nuclear Instruments and Methods A .

\bibitem{Dickinson:2000}
J.~Dickinson et~al.,
\newblock Nuclear Instruments and Methods A {\bf 440}, 95 (2000).

\bibitem{Ahrens:2002}
J.~Ahrens et~al.,
\newblock Physical Review D {\bf 66}, 012005 (2002).

\bibitem{Abbasi:2009}
R.~Abbasi et~al.,
\newblock Nuclear Instruments and Methods A {\bf 601}, 294 (2008).

\bibitem{Achterberg:2006}
A.~Achterberg et~al.,
\newblock Astroparticle Physics {\bf 26}, 155 (2006).

\bibitem{Andeen:thesis}
K.~G. Andeen,
\newblock {\em First Measurements of Cosmic Ray Composition from 1-50 PeV Using
  New Techniques on Coincident Data from the IceCube Neutrino Observatory},
\newblock PhD thesis, University of Wisconsin-Madison, 2011.

\bibitem{Heck:2010}
D.~Heck et~al.,
\newblock {\em Forschungszentrum Karlsruhe Report FZKA 6019}, 1998.

\bibitem{Ahn:2009}
E.~J. Ahn, R.~Engel, T.~K. Gaisser, P.~Lipari, and T.~Stanev,
\newblock Physical Review D {\bf 80}, 94003 (2009).

\bibitem{FLUKA:2006}
G.~Battistoni et~al.,
\newblock AIP Conference Proceedings {\bf 896}, 31 (2007).

\bibitem{GEANT4:2003}
S.~Agostinelli et~al.,
\newblock Nuclear Instruments and Methods A {\bf 506}, 250 (2003).

\bibitem{GEANT4:2006}
J.~Allison et~al.,
\newblock IEEE Transactions on Nuclear Science {\bf 53}, 270 (2006).

\bibitem{Chirkin:2004}
D.~Chirkin and W.~Rhode,
\newblock preprint {\bf arXiv:hep-ph/0407075} (2004).

\bibitem{Lundberg:2007}
J.~Lundberg,
\newblock Nuclear Instruments and Methods {\bf A581}, 619 (2007).

\bibitem{IceModel:2006}
M.~Ackermann et~al.,
\newblock J. Geophys. Res. {\bf 111}, D13203 (2006).

\bibitem{Klepser:thesis}
S.~Klepser,
\newblock {\em Reconstruction of Extensive Air Showers and Measurement of the
  Cosmic Ray Energy Spectrum in the Range of 1-80 PeV at the South Pole},
\newblock PhD thesis, Humbolt-Universit\"{a}t zu Berlin, 2008.

\bibitem{Kislat:2011}
R.~Abbasi et~al.,
\newblock submitted to Astroparticle Physics {\bf arXiv:1202.3039v1} (2012).

\bibitem{Ahrens:2004a}
J.~Ahrens et~al.,
\newblock Nuclear Instruments and Methods A {\bf 524}, 169 (2004).

\bibitem{Rawlins:thesis}
K.~Rawlins,
\newblock {\em Measuring the Composition of Cosmic Rays with the SPASE and
  AMANDA Detectors},
\newblock PhD thesis, University of Wisconsin-Madison, 2001.

\bibitem{Ahrens:2004}
J.~Ahrens et~al.,
\newblock Astroparticle Physics {\bf 21}, 565 (2004).

\bibitem{Feusels:2009}
T.~Feusels, J.~Eisch, and C.~Xu,
\newblock Proceedings of the 31st International Cosmic Ray Conference, Lodz,
  Poland {\bf arXiv:0912.4668v1} (2009).

\bibitem{Bishop:1994}
C.~M. Bishop,
\newblock Review of Scientific Instruments {\bf 65}, 1803 (1994).

\bibitem{ROOT}
The ROOT Team at CERN,
\newblock {\em ROOT: An Object-Oriented Data Analysis Framework, Users Guide}.

\bibitem{Horandel:2003}
J.~H\"{o}randel,
\newblock Astroparticle Physics {\bf 19}, 193 (2003).

\bibitem{Antonyan:2000}
S.~V. Ter-Antonyan and L.~S. Haroyan,
\newblock preprint {\bf arXiv:hep-ex/0003006v3} (2000).

\bibitem{MINUIT:1975}
F.~James et~al.,
\newblock Computer Physics Communications {\bf 10}, 343 (1975).

\bibitem{EPOS:2007}
K.~Werner,
\newblock Nuclear Physics B Proceedings Supplements {\bf 175}, 81 (2007).

\bibitem{qgsjet:1997}
N.~N. Kalmykov, S.~S. Ostapchenko, and A.~I. Pavlov,
\newblock Nuclear Physics B Proceedings Supplements {\bf 52}, 17 (1997).

\bibitem{Chirkin:2011}
D.~Chirkin,
\newblock Proceedings of the 32nd International Cosmic Ray Conference, Beijing,
  China {\bf 4}, 161 (2011).

\bibitem{Price:2000}
P.~B. Price, K.~Woschnagg, and D.~Chirkin,
\newblock Geophys. Res. Letters {\bf 27}, 2129 (2000).

\bibitem{Abbasi:2010a}
R.~Abbasi et~al.,
\newblock Nuclear Instruments and Methods in Physics Research Section A {\bf
  618}, 139 (2010).

\bibitem{Kampert:2012}
K.-H. Kampert and M.~Unger,
\newblock Astroparticle Physics {\bf 35}, 660 (2012).

\end{thebibliography}

\end{document}